\documentclass[11pt]{article}
\usepackage{epsfig} 
\newcommand{\sect}[1]{ \section{#1} \setcounter{equation}{0} }

\newcommand{\pslash}{p \! \! \! /} 
\newcommand{\qslash}{q \! \! \! /}

\newcommand{\half}{\mbox{\small{$\frac{1}{2}$}}} 
\newcommand{\third}{\mbox{\small{$\frac{1}{3}$}}} 
 
\newcommand{\pitwo}{\mbox{\small{$\frac{\pi}{2}$}}} 
\newcommand{\pisix}{\mbox{\small{$\frac{\pi}{6}$}}} 
\newcommand{\MSbar}{\overline{\mbox{MS}}} 
\newcommand{\MSbars}{\overline{\mbox{\footnotesize{MS}}}} 
\newcommand{\Nf}{N_{\!f}}
\newcommand{\NF}{N_{\!F}}

\begin{document}
\title{Two loop renormalization of the $n$~$=$~$2$ Wilson operator in the
RI${}^\prime$/SMOM scheme}
\author{J.A. Gracey, \\ Theoretical Physics Division, \\ 
Department of Mathematical Sciences, \\ University of Liverpool, \\ P.O. Box 
147, \\ Liverpool, \\ L69 3BX, \\ United Kingdom.} 
\date{} 
\maketitle 

\vspace{5cm} 
\noindent 
{\bf Abstract.} We compute the anomalous dimensions of the flavour non-singlet
twist-$2$ Wilson operators in the RI${}^\prime$/SMOM scheme at two loops in an
arbitrary linear covariant gauge. In addition we provide the full Green's 
function for these operators inserted in a quark $2$-point function at the 
symmetric subtraction point. The three loop anomalous dimensions in the Landau
gauge are also derived. 

\vspace{-16.2cm}
\hspace{13.5cm}
{\bf LTH 910}

\newpage

\sect{Introduction.}

In a quantum field theory the behaviour of the Green's functions or $n$-point
functions derived from the Lagrangian carry all the information about the
dynamics of the quantum particles. For the vast majority of quantum field
theories, however, it is impossible to extract their behaviour for all ranges
of momenta and parameters, such as the particle masses and coupling constants.
Instead one invariably examines them in various regions of interest, such as at
high or low energy, using a variety of techniques. For instance, at high
energy in Quantum Chromodynamics (QCD) one uses perturbation theory since the
coupling constant is small as a consequence of asymptotic freedom, \cite{1,2}.
By contrast at low energy perturbation theory breaks down and non-perturbative
methods have been developed and refined to give credible information. The
central tool at this energy is lattice gauge theory involving an intense
amount of numerical computations on high performance computers. This approach 
has in general been hugely successful in determining bound state masses, for 
example, and exploring the structure of nucleons. For instance, matrix elements
of the underlying operators used in deep inelastic scattering, known as Wilson 
operators, play a key role in this, \cite{3}. The behaviour of such matrix 
elements at low energy is useful in extracting theoretical information for 
parton structure functions. In outlining the general aspects of Green's 
functions in understanding the dynamics of the strong nuclear force, we are 
overlooking the huge technical effort which is required to ensure accurate 
estimates are obtained. For instance, as one is dealing with a quantum field 
theory, the operators undergo renormalization, \cite{3}. Equally when one 
produces estimates from a low energy computation on the lattice one has to be 
assured that the result is consistent with and extrapolates onto the high 
energy behaviour of the same object which can be computed within perturbation 
theory. Indeed there has been a large degree of activity on the lattice in this
respect for quark currents and Wilson operators. For instance, see 
\cite{4,5,6,7,8,9,10,11,12,13,14,15,16,17,18,19,20} for representative
analyses. 

For the continuum computations one usually calculates in the $\MSbar$ scheme 
which is a mass independent renormalization scheme. The advantage of this 
scheme is that it is one in which the largest order of perturbation theory can 
be determined. However, defining the same scheme for lattice calculations is 
not as easy since it invariably requires a numerical differentiation on the 
lattice. Taking such derivatives carries a financial penalty. Therefore, 
alternative schemes have been developed for lattice gauge theory which avoids 
the use of derivatives. Such a class of schemes is generally referred to as 
Regularization Invariant (RI), \cite{21,22}. However, there are two main types.
The original class involves RI itself and a modified version known as 
RI${}^\prime$, \cite{21,22}. These are similar in that QCD is renormalized for 
$3$-point and higher Green's functions according to the $\MSbar$ prescription
but for $2$-point functions such as those determining propagator corrections, 
the renormalization condition is defined by ensuring that the contributions
from radiative corrections at the subtraction point are absent. In this class 
of schemes we include the zero momentum insertion of an operator in, say, a 
quark $2$-point function. The modified scheme, RI${}^\prime$, differs from RI 
in the way the quark wave function is defined, \cite{21,22}. Whilst originally 
defined in \cite{21,22} specifically in the context of the lattice, this scheme
has also been studied in the Landau gauge in the continuum at three and four 
loops, \cite{23}, and at three loops in an arbitrary linear covariant gauge, 
\cite{24}. Subsequently, the Green's functions of a variety of low moment 
twist-$2$ flavour non-singlet operators central to deep inelastic scattering 
inserted in a quark $2$-point function were evaluated to three loops in 
$\MSbar$ and RI${}^\prime$, \cite{25,26}. This high order of perturbation 
theory provided useful information on matching the lattice measurement of the 
same object at high energy. 

More recently a second class of regularization invariant schemes has been
developed in \cite{27,28,29}. It is termed RI${}^\prime$/SMOM where the first 
part of the designation indicates the RI${}^\prime$ scheme definition of the 
quark wave function renormalization. The second refers to the method used to 
define the renormalization constants of $3$-point functions with an operator 
insertion at non-zero momentum. As there are momenta flowing into each point of
the Green's function, the renormalization is carried out at a symmetric point
where the squared momenta of all three incoming momenta take the {\em same}
value which is the origin of the S. The MOM indicates the ethos that was 
present in the overall RI definition in that the condition for defining the 
operator renormalization constant is to ensure that the part corresponding to
radiative corrections at this symmetric subtraction point are absent. This 
particular scheme was developed to avoid the strong sensitivity of the 
RI${}^\prime$ scheme to infrared effects, \cite{27}, as well as to try and have
a more rapidly convergent perturbation series for the conversion functions to 
the $\MSbar$ scheme. It was initially applied to the renormalization of the 
quark currents at one loop in \cite{27} to determine the anomalous dimensions, 
amplitudes and conversion functions in the $\MSbar$ scheme. Subsequently, the 
two loop renormalization of the scalar and tensor currents was treated in 
\cite{28,29} with the anomalous dimensions and conversion functions being
computed. Though with the knowledge of the two loop conversion functions the 
three loop Landau gauge anomalous dimensions were deduced too. More recently, 
the full set of amplitudes for the scalar, vector and tensor currents have been
calculated at two loops in \cite{30}. 

Given that the measurement of nucleon matrix elements is of interest it is the 
purpose of this article to provide the anomalous dimensions and amplitudes for 
the second moment of the flavour non-singlet Wilson operator at two loops in 
RI${}^\prime$/SMOM. This will build on the analogous one loop computation of 
\cite{31}. The treatment of this operator is complicated by the fact that it 
mixes with a total derivative operator. Ordinarily one is not concerned by this
extra operator since if it were inserted at zero momentum it would give no 
contribution to the Green's function. However, the symmetric subtraction point 
of the RI${}^\prime$/SMOM scheme means that such total derivative operators 
play an active role and cannot be ignored. The full three loop $\MSbar$ mixing 
matrix of anomalous dimensions for this second moment operator was given in 
\cite{32}. With the two loop conversion functions we will provide the three 
loop result in the RI${}^\prime$/SMOM scheme. One feature of the 
RI${}^\prime$/SMOM construction for the Wilson operators is the role the vector
current has in the renormalization. Its relation to the Slavnov-Taylor identity
was discussed in \cite{27} at one loop, as well as two loop, \cite{30}, and in  
the context of the Wilson operators in \cite{31}. The divergence of the vector 
current resides within the total derivative operator into which the Wilson 
operator mixes. Hence, the two loop renormalization of the vector current and 
the associated RI${}^\prime$/SMOM amplitudes of the Green's function where this
was determined, \cite{30}, are important as checks on the computation presented 
here. Indeed partly related to this is that one has to ensure the 
renormalization of the second moment is consistent with one of the 
Slavnov-Taylor identities of QCD. Further, the third moment of the Wilson 
operator was also considered at one loop in \cite{31}. From that it transpires
that the renormalization of the higher moment operators is dependent on that of
the lower moment operators of the tower with the vector current at the 
foundation. Therefore, the treatment of the $n$~$=$~$2$ moment is relevant for 
that of the higher moment operators. Throughout we will use $\Nf$ flavours of 
massless quarks and therefore all our expressions are in the chiral limit. 
Including masses for quarks is not currently viable as the basic scalar master 
Feynman integrals have not been evaluated for the momentum configuration of the
Green's function we consider here. 

The article is organized as follows. The background to the formalism we will 
use, notation and method to define and extract the scalar amplitudes of 
interest are discussed in section two. Section three is devoted to summarizing 
the main results of the RI${}^\prime$/SMOM computation of the second moment of 
the Wilson operator. The amplitudes are given numerically. Though the full 
exact expressions are recorded in the form of a set of tables. The results 
relating to the actual operator renormalization, such as the anomalous 
dimensions and conversion functions are provided in section four whilst we
summarize our conclusions in section five.

\sect{Preliminaries.}

First, we recall the key properties of the operators we will consider. We use
the notation of \cite{32} in this respect and denote
\begin{equation}
W_2 ~ \equiv ~ {\cal S} \bar{\psi} \gamma^\mu D^\nu \psi ~~~,~~~ 
\partial W_2 ~ \equiv ~ {\cal S} \partial^\mu \left( \bar{\psi} \gamma^{\nu}
\psi \right) 
\end{equation}
where $D_\mu$ is the covariant derivative with all derivatives acting to the 
right and ${\cal S}$ indicates that the operator is traceless and symmetric in 
its Lorentz indices. Moreover, as we will be using dimensional regularization 
we regard the operators as being traceless in $d$-dimensions. So, for instance,
we have 
\begin{equation}
{\cal S} {\cal O}^{W_2}_{\mu\nu} ~=~ {\cal O}^{W_2}_{\mu\nu} ~+~
{\cal O}^{W_2}_{\nu\mu} ~-~ \frac{2}{d} \eta_{\mu\nu}
{\cal O}^{W_2\,\sigma}_{\sigma} 
\end{equation}
where
\begin{equation}
{\cal O}^{W_2}_{\mu\nu} ~=~ \bar{\psi} \gamma_\mu D_\nu \psi ~. 
\end{equation}
The total derivative operator denoted by $\partial W_2$ also obeys the same
symmetrization. Throughout we concentrate on flavour non-singlet operators and
omit the flavour indices on the quark fields $\psi$. Ordinarily when one
renormalizes $W_2$ in perturbation theory the operator is inserted in a quark
$2$-point function, \cite{3,33,34}, at zero momentum. Then the renormalization 
constant emerges under the assumption that the renormalization of this operator
is multiplicative. This is not strictly speaking true. It is only true for that
particular momentum routing through the quark $2$-point Green's function with 
the operator insertion. If instead there was a momentum flowing out through the
operator then there is a mixing with the total derivative operator. The Feynman
rule for the latter operator vanishes when there is no momentum flow out 
through the operator which is why the momentum configuration of \cite{3,33,34}
is multiplicative. In other words for the full renormalization of $W_2$ one has
to determine the mixing matrix of renormalization constants where 
$\partial W_2$ completes the set. As we are considering the renormalization of 
$W_2$ in the RI${}^\prime$/SMOM scheme we have to take this into account as the
momentum setup for the underlying Green's function involves two independent 
momenta flowing in through the external quark legs and out through the operator
itself. This is illustrated in Figure $1$. 

\begin{figure}[ht]
\hspace{6cm}
\epsfig{file=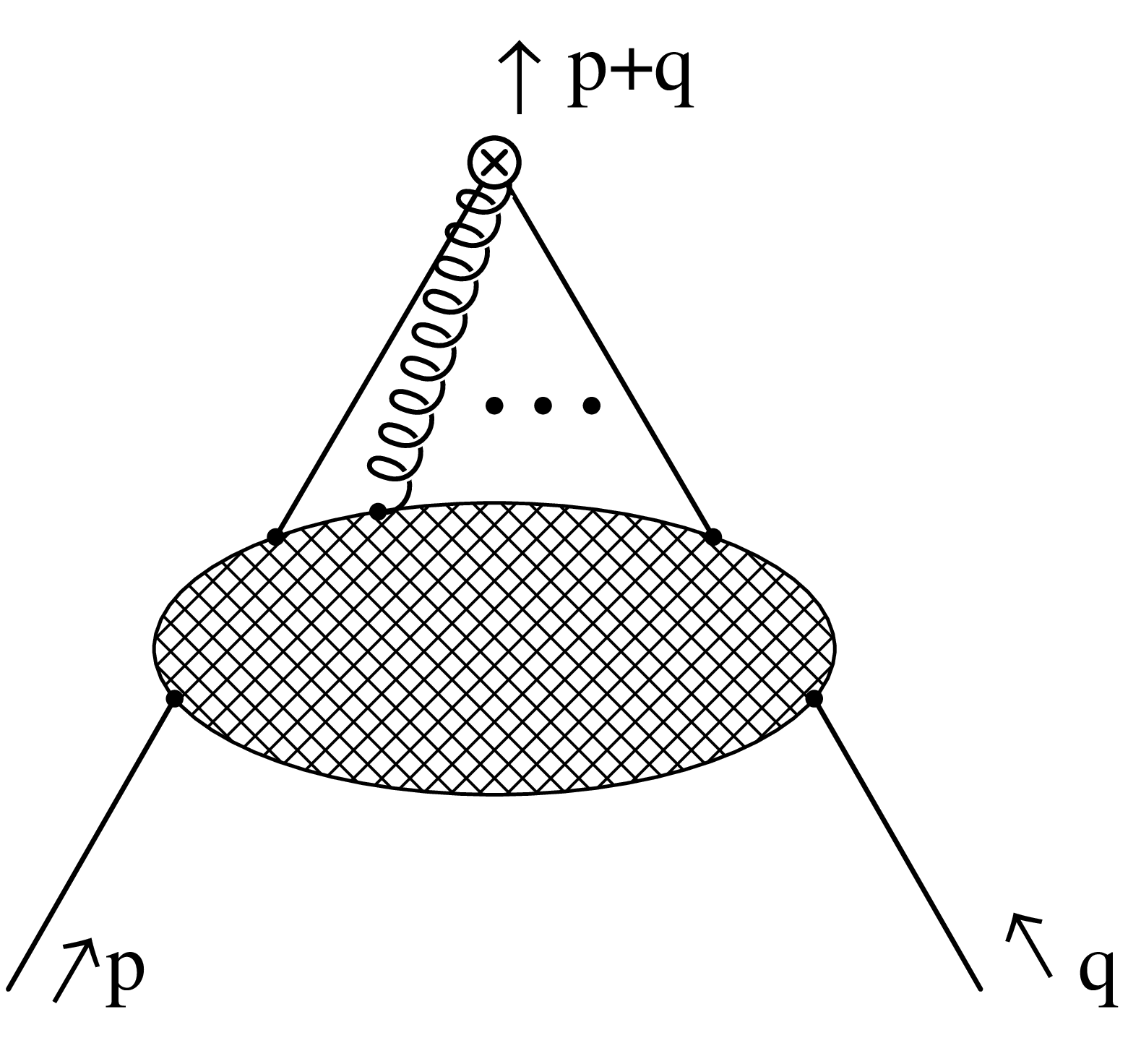,height=5cm}
\vspace{0.5cm}
\caption{The Green's function $\left\langle \psi(p) {\cal O}^i_{\mu_1 \ldots 
\mu_{n_i}} (-p-q) \bar{\psi}(q) \right\rangle$.}
\end{figure}

The renormalization of the set of level $W_2$ operators has been carried out to
three loops in the $\MSbar$ scheme in \cite{32}. As we will require these later
for the three loop RI${}^\prime$/SMOM scheme anomalous dimensions later we
briefly recall the relevant results and formalism. First, irrespective of which
scheme one works in the mixing matrix of renormalization constants is 
\begin{equation}
Z^{W_2}_{ij} ~=~ \left(
\begin{array}{cc}
Z^{W_2}_{11} & Z^{W_2}_{12} \\
0 & Z^{W_2}_{22} \\
\end{array}
\right) ~.
\label{mixmat}
\end{equation}
Here we retain the notation of \cite{31,32} whereby the superscript indicates
the operator level $W_2$ and the labels indicating the matrix elements 
respectively label the two elements of the operators in that set which are 
$W_2$ and $\partial W_2$. Whilst $W_2$ is used to mean level and operator it 
will be clear from the context which is meant. The anomalous dimensions are 
formally defined by  
\begin{equation}
\gamma^{W_2}_{ij} (a,\alpha) ~=~ -~ \mu \frac{d ~}{d \mu} \ln Z^{W_2}_{ij}
\end{equation}
where here and below $W_2$ represents the level, from which one can deduce the 
relations between the individual matrix elements are
\begin{eqnarray}
0 &=& \gamma^{W_2}_{11}(a,\alpha) Z^{W_2}_{11} ~+~
\mu \frac{d ~}{d \mu} Z^{W_2}_{11} \nonumber \\
0 &=& \gamma^{W_2}_{11}(a,\alpha) Z^{W_2}_{12} ~+~
\gamma^{W_2}_{12}(a,\alpha) Z^{W_2}_{22} ~+~
\mu \frac{d ~}{d \mu} Z^{W_2}_{12} \nonumber \\
0 &=& \gamma^{W_2}_{22}(a,\alpha) Z^{W_2}_{22} ~+~
\mu \frac{d ~}{d \mu} Z^{W_2}_{22} 
\end{eqnarray}
with
\begin{equation}
\mu \frac{d~}{d\mu} ~=~ \beta(a) \frac{\partial ~}{\partial a} ~+~
\alpha \gamma_\alpha(a,\alpha) \frac{\partial ~}{\partial \alpha} 
\end{equation}
and $\gamma_\alpha(a,\alpha)$ is the anomalous dimension of the gauge
parameter. We retain our previous conventions, \cite{31,32}, for consistency. 
In the $\MSbar$ scheme since the operators $W_2$ and $\partial W_2$ are gauge 
invariant then the anomalous dimensions are independent of the gauge parameter.
However, for mass dependent renormalization schemes, of which 
RI${}^\prime$/SMOM is an example, the anomalous dimensions are $\alpha$ 
dependent. Therefore, we recall that for the $\MSbar$ scheme
\begin{eqnarray}
\left. \frac{}{} \gamma^{W_2}_{11}(a) \right|_{\MSbars} &=& 
\frac{8}{3} C_F a ~+~ \frac{1}{27} \left[ 376 C_A C_F - 112 C_F^2 
- 128 C_F T_F \Nf \right] a^2 \nonumber \\
&& +~ \frac{1}{243} \left[ \left( 5184 \zeta(3) + 20920 \right) C_A^2 C_F
- \left( 15552 \zeta(3) + 8528 \right) C_A C_F^2 \right. \nonumber \\
&& \left. ~~~~~~~~~~-~ \left( 10368 \zeta(3) + 6256 \right) C_A C_F T_F \Nf
+ \left( 10368 \zeta(3) - 560 \right) C_F^3 \right. \nonumber \\
&& \left. ~~~~~~~~~~+~ \left( 10368 \zeta(3) - 6824 \right) C_F^2 T_F \Nf
- 896 C_F T_F^2 \Nf^2 \right] a^3 ~+~ O(a^4) \nonumber \\
\left. \frac{}{} \gamma^{W_2}_{12}(a) \right|_{\MSbars} &=& 
-~ \frac{4}{3} C_F a ~+~ \frac{1}{27} \left[ 56 C_F^2 - 188 C_A C_F 
+ 64 C_F T_F \Nf \right] a^2 \nonumber \\
&& +~ \frac{1}{243} \left[ \left( 7776 \zeta(3) + 4264 \right) C_A C_F^2
- \left( 2592 \zeta(3) + 10460 \right) C_A^2 C_F \right. \nonumber \\
&& \left. ~~~~~~~~~~+~ \left( 5184 \zeta(3) + 3128 \right) C_A C_F T_F \Nf
- \left( 5184 \zeta(3) - 280 \right) C_F^3 \right. \nonumber \\
&& \left. ~~~~~~~~~~-~ \left( 5184 \zeta(3) - 3412 \right) C_F^2 T_F \Nf
+ 448 C_F T_F^2 \Nf^2 \right] a^3 ~+~ O(a^4) \nonumber \\
\left. \frac{}{} \gamma^{W_2}_{22}(a) \right|_{\MSbars} &=& O(a^4)
\label{anomdimms}
\end{eqnarray}
where $\zeta(z)$ is the Riemann zeta function and for this scheme we have 
omitted the part of the argument involving $\alpha$ to emphasise the absence of
gauge dependence. The colour group Casimirs are defined by
\begin{equation}
T^a T^a ~=~ C_F ~~~,~~~ f^{acd} f^{bcd} ~=~ C_A \delta^{ab} ~~~,~~~ 
\mbox{Tr} \left( T^a T^b \right) ~=~ T_F \delta^{ab}
\end{equation}
where $T^a$ are the colour group generators with structure constants $f^{abc}$
and $1$~$\leq$~$a$~$\leq$~$N_A$ with $N_A$ the dimension of the adjoint
representation. The coupling constant appearing in the covariant derivative is
$g$ but appears in (\ref{anomdimms}) in the combination 
$a$~$=$~$g^2/(16\pi^2)$. The absence of a correction to $\left. \!\!\! 
\frac{}{} \gamma^{W_2}_{22}(a) \right|_{\MSbars}$ to this order is a reflection
of the all orders result where there is no renormalization of $\partial W_2$.
This is because it corresponds to the vector current which is a physical 
operator and is not renormalized reflecting the Slavnov-Taylor identity of the 
underlying gauge theory. See, for example, \cite{35} for background. Moreover, 
this property is retained in all schemes and we must ensure that the 
RI${}^\prime$/SMOM scheme renormalization respects this general feature. 

In order to define the RI${}^\prime$/SMOM renormalization scheme for level
$W_2$ we compute the Green's function illustrated in Figure $1$ where the
external momenta $p$ and $q$ are independent. However, we will focus totally on
its structure at the symmetric subtraction point, \cite{27}, which is defined
as 
\begin{equation}
p^2 ~=~ q^2 ~=~ ( p + q )^2 ~=~ -~ \mu^2 
\end{equation}
where $\mu$ is the common mass scale. These relations imply 
\begin{equation}
pq ~=~ \frac{1}{2} \mu^2 ~.
\end{equation}
In order to determine the renormalization constant in the $\MSbar$ or
RI${}^\prime$ schemes one can proceed with only one external momentum. For
instance, the choice $p$~$=$~$-$~$q$ corresponds to a zero momentum insertion
{\em before} setting $p^2$ to be a specific scale involving $\mu^2$. Though in
fact to determine the full mixing matrix in the $\MSbar$ scheme requires 
accessing the total derivative operators which is achieved by nullifiying the 
momentum of one of the external quark legs, \cite{32}. As the operators we 
consider have free Lorentz indices then the Green's function has to be 
decomposed into a set of Lorentz invariant amplitudes. The basis for $W_2$ has 
already been constructed in \cite{31} and we briefly recall the construction as
well as the notation. First, we write the Green's function as 
\begin{equation}
\left. \left\langle \psi(p) {\cal O}^i_{\mu_1 \ldots \mu_{n_i}}(-p-q) 
\bar{\psi}(q) \right\rangle \right|_{p^2 = q^2 = - \mu^2} ~=~ \sum_{k=1}^{n_i} 
{\cal P}^i_{(k) \, \mu_1 \ldots \mu_{n_i} }(p,q) \, 
\Sigma^{{\cal O}^i}_{(k)}(p,q) ~.
\end{equation}
Here $\Sigma^{{\cal O}^i}_{(k)}(p,q)$ are the scalar amplitudes for each of the
two operators ${\cal O}^i$ whereas ${\cal P}^i_{(k) \, \mu_1 \ldots 
\mu_{n_i} }(p,q)$ are the Lorentz tensors of the basis whose Lorentz indices 
reflect the symmetries of the operator which has been inserted. For level $W_2$
there are ten independent Lorentz tensors and we will detail these later. To
evaluate the scalar amplitudes we use the method of projection where we apply
different linear combinations of the basis tensors to the Green's function in 
such a way that each amplitude is isolated in turn. More specifically we have,
\cite{32},
\begin{equation}
\Sigma^{{\cal O}^i}_{(k)}(p,q) ~=~ {\cal M}^i_{kl} 
{\cal P}^{i ~\, \mu_1 \ldots \mu_{n_i}}_{(l)}(p,q) \left. \left(  
\left\langle \psi(p) {\cal O}^i_{\mu_1 \ldots \mu_{n_i}}(-p-q) \bar{\psi}(q) 
\right\rangle \right) \right|_{p^2=q^2=-\mu^2} 
\end{equation}
where ${\cal M}^i_{kl}$ is a matrix of rational polynomials in the spacetime
dimension $d$. It is constructed by first determining the matrix 
${\cal N}^i_{kl}$ which is defined by the contraction of the tensors of the 
basis with each of the other tensors via 
\begin{equation}
{\cal N}^i_{kl} ~=~ \left. {\cal P}^i_{(k) \, \mu_1 \ldots \mu_{n_i}}(p,q)
{\cal P}^{i ~\, \mu_1 \ldots \mu_{n_i}}_{(l)}(p,q) \right|_{p^2=q^2=-\mu^2} ~. 
\end{equation}
Then ${\cal M}^i_{kl}$ is the inverse of ${\cal N}^i_{kl}$.

More specifically, the basis of Lorentz tensors into which we decompose the 
Green's functions with the operator insertions is, \cite{32}, 
\begin{eqnarray}
{\cal P}^{W_2}_{(1) \mu \nu }(p,q) &=& \gamma_\mu p_\nu + \gamma_\nu p_\mu
- \frac{2}{d} \pslash \eta_{\mu\nu} ~~,~~
{\cal P}^{W_2}_{(2) \mu \nu }(p,q) ~=~ \gamma_\mu q_\nu + \gamma_\nu q_\mu
- \frac{2}{d} \qslash \eta_{\mu\nu} \nonumber \\
{\cal P}^{W_2}_{(3) \mu \nu }(p,q) &=& \pslash \left[
\frac{1}{\mu^2} p_\mu p_\nu + \frac{1}{d} \eta_{\mu\nu} \right] ~~,~~
{\cal P}^{W_2}_{(4) \mu \nu }(p,q) ~=~ \pslash \left[
\frac{1}{\mu^2} p_\mu q_\nu + \frac{1}{\mu^2} q_\mu p_\nu
- \frac{1}{d} \eta_{\mu\nu} \right] \nonumber \\
{\cal P}^{W_2}_{(5) \mu \nu }(p,q) &=& \pslash \left[
\frac{1}{\mu^2} q_\mu q_\nu + \frac{1}{d} \eta_{\mu\nu} \right] ~~,~~
{\cal P}^{W_2}_{(6) \mu \nu }(p,q) ~=~ \qslash \left[
\frac{1}{\mu^2} p_\mu p_\nu + \frac{1}{d} \eta_{\mu\nu} \right] \nonumber \\
{\cal P}^{W_2}_{(7) \mu \nu }(p,q) &=& \qslash \left[
\frac{1}{\mu^2} p_\mu q_\nu + \frac{1}{\mu^2} q_\mu p_\nu
-\frac{1}{d} \eta_{\mu\nu} \right] ~,~
{\cal P}^{W_2}_{(8) \mu \nu }(p,q) ~=~ \qslash \left[
\frac{1}{\mu^2} q_\mu q_\nu + \frac{1}{d} \eta_{\mu\nu} \right] \nonumber \\
{\cal P}^{W_2}_{(9) \mu \nu }(p,q) &=& \frac{1}{\mu^2} \left[
\Gamma_{(3) \, \mu p q } p_\nu + \Gamma_{(3) \, \nu p q } p_\mu \right]
\nonumber \\ 
{\cal P}^{W_2}_{(10) \mu \nu }(p,q) &=& \frac{1}{\mu^2} \left[
\Gamma_{(3) \, \mu p q } q_\nu + \Gamma_{(3) \, \nu p q } q_\mu \right] ~.
\label{basten}
\end{eqnarray}
As there are ten elements in the basis then in order to save space for 
presenting the projection matrix, we have partitioned the $10$~$\times$~$10$ 
matrix into four sub-matrices. Defining 
\begin{eqnarray}
{\cal M}^{W_2} &=& -~ \frac{1}{108(d-2)} \left(
\begin{array}{cc}
{\cal M}^{W_2}_{11} & {\cal M}^{W_2}_{12} \\ 
{\cal M}^{W_2}_{21} & {\cal M}^{W_2}_{22} \\ 
\end{array}
\right)
\end{eqnarray}
then we have
\begin{eqnarray}
{\cal M}^{W_2}_{11} &=& \left(
\begin{array}{ccccc}
18 & 9 & 48 & 24 & 12 \\
9 & 18 & 24 & 30 & 24 \\
48 & 24 & 64 (d+1) & 32 (d+1) & 16 (d+4) \\
24 & 30 & 32 (d+1) & 8 (5d-1) & 8 (4d+1) \\
12 & 24 & 16 (d+4) & 8 (4d+1) & 32 (2d-1) \\
\end{array}
\right) ~, \nonumber \\
{\cal M}^{W_2}_{12} &=& \left(
\begin{array}{ccccc}
24 & 30 & 24 & 0 & 0 \\
12 & 24 & 48 & 0 & 0 \\
32 (d+1) & 16 (d+4) & 8 (d+10) & 0 & 0 \\
16 (d+1) & 20 (d+1) & 16 (d+4) & 0 & 0 \\
8 (d+4) & 16 (d+1) & 32 (d+1) & 0 & 0 \\
\end{array}
\right) ~, \nonumber \\
{\cal M}^{W_2}_{21} &=& \left(
\begin{array}{ccccc}
24 & 12 & 32 (d+1) & 16 (d+1) & 8 (d+4) \\
30 & 24 & 16 (d+4) & 20 (d+1) & 16 (d+1) \\
24 & 48 & 8 (d+10) & 16 (d+4) & 32 (d+1) \\
0 & 0 & 0 & 0 & 0 \\
0 & 0 & 0 & 0 & 0 \\
\end{array}
\right) ~, \nonumber \\
{\cal M}^{W_2}_{22} &=& \left(
\begin{array}{ccccc}
32 (2d-1) & 8 (4d+1) & 16 (d+4) & 0 & 0 \\
8 (4d+1) & 8 (5d-1) & 32 (d+1) & 0 & 0 \\
16 (d+4) & 32 (d+1) & 64 (d+1) & 0 & 0 \\
0 & 0 & 0 & - 24 & - 12 \\
0 & 0 & 0 & - 12 & - 24 \\
\end{array}
\right) ~.
\end{eqnarray}
We note that here the label $W_2$ refers to the level and the same projection 
is used for $\partial W_2$. In defining this basis we have used the set of
generalized $\gamma$-matrices denoted by $\Gamma_{(n)}^{\mu_1 \ldots \mu_n}$
which are totally antisymmetric in the Lorentz indices and defined by, 
\cite{36,37,38}, 
\begin{equation}
\Gamma_{(n)}^{\mu_1 \ldots \mu_n} ~=~ \gamma^{[\mu_1} \ldots \gamma^{\mu_n]}
\label{gamdef}
\end{equation}
where the notation includes the overall factor of $1/n!$. For simplicity we 
will retain the more natural notation for a single $\gamma$-matrix rather than
use the clumsy $\Gamma_{(1)}^\mu$. General properties have already been 
discussed in \cite{39,40} but we note that one of particular use is 
\begin{equation}
\mbox{tr} \left( \Gamma_{(m)}^{\mu_1 \ldots \mu_m}
\Gamma_{(n)}^{\nu_1 \ldots \nu_n} \right) ~ \propto ~ \delta_{mn}
I^{\mu_1 \ldots \mu_m \nu_1 \ldots \nu_n}
\end{equation}
where $I^{\mu_1 \ldots \mu_m \nu_1 \ldots \nu_n}$ is the unit matrix in the
space of these generalized $\gamma$-matrices. This leads to the natural 
partition of ${\cal M}^{W_2}$ into two submatrices. The use of  
$\Gamma_{(n)}^{\mu_1 \ldots \mu_n}$ is especially appropriate as we will be
using dimensional regularization and these objects form the complete basis for 
spinor space in $d$-dimensional spacetimes. Moreover, they allow us to see that
the set (\ref{basten}) is complete. If there were more than two independent 
momenta in this problem then one would have to include matrices with 
$n$~$\geq$~$4$. One final comment on the basis choice and that is that the
basis is not unique. One could choose different linear combinations of the
tensors we use here but the overall structure of the Green's function would be
unaltered. Though we have chosen to retain a degree of symmetry if one
interchanges $p$ and $q$. For the total derivative operator, $\partial W_2$,
which is symmetric under this then the amplitudes will also respect this
property as will be evident in the explicit expressions. 

Having dealt with the general structure of $\left. \left\langle \psi(p) 
{\cal O}^i_{\mu_1 \ldots \mu_{n_i}}(-p-q) \bar{\psi}(q) \right\rangle 
\right|_{p^2 = q^2 = - \mu^2}$ we now need to outline the procedure to 
renormalize our operators in the RI${}^\prime$/SMOM scheme. There are a variety
of ways of doing this. If for the moment we denote by $0$ a particular
amplitude or combination of amplitudes which contains all the divergent parts
of the Green's function then we will use the renormalization condition 
\begin{equation}
\left. \lim_{\epsilon\rightarrow 0} \left[ 
Z^{\mbox{\footnotesize{RI$^\prime$}}}_\psi 
Z^{\mbox{\footnotesize{RI$^\prime$/SMOM}}}_{\cal O} \Sigma^{\cal O}_{(0)}(p,q) 
\right] \right|_{p^2 \, = \, q^2 \, = \, - \mu^2} ~=~ 1 
\end{equation} 
where $d$~$=$~$4$~$-$~$2\epsilon$ in dimensional regularization. This is in 
keeping with the overall regularization invariant scheme ethos that there 
should be no $O(a)$ finite parts for this particular amplitude combination $0$,
\cite{27,28,29}. The renormalization constant 
$Z^{\mbox{\footnotesize{RI$^\prime$/SMOM}}}_{\cal O}$ is the one which removes
the divergences and leads to the anomalous dimension of the operator. However,
as the operator has been inserted into a quark $2$-point function then the wave
function renormalization constant of these external quarks has to be considered
which is reflected in $Z^{\mbox{\footnotesize{RI$^\prime$}}}_\psi$. The 
annotation here is RI${}^\prime$ as the $2$-point functions of the theory are 
renormalized according to the original procedure of \cite{21,22,23}. In the
RI${}^\prime$ scheme the $3$-point and higher functions are renormalized
according to the $\MSbar$ scheme so that their finite parts after 
renormalization are not unity. In \cite{23} the Landau gauge three and four
loop RI and RI${}^\prime$ scheme quark field and mass anomalous dimensions were
recorded. The anomalous dimensions of all the fields in an arbitrary linear 
covariant gauge were given at three loops in \cite{24}. In the latter article 
the renormalization of the gauge parameter was also determined and the relation
between the parameter definition in different schemes was established. For 
instance, if we denote the scheme in which the variable is defined in by a 
subscript then
\begin{equation}
a_{\mbox{\footnotesize{RI$^\prime$}}} ~=~
a_{\mbox{\footnotesize{$\MSbars$}}} ~+~ O \left(
a_{\mbox{\footnotesize{$\MSbars$}}}^5 \right)
\end{equation}
and, \cite{24}, 
\begin{eqnarray}
\alpha_{\mbox{\footnotesize{RI$^\prime$}}}
&=& \left[ 1 + \left( \left( - 9 \alpha_{\mbox{\footnotesize{$\MSbars$}}}^2
- 18 \alpha_{\mbox{\footnotesize{$\MSbars$}}} - 97 \right) C_A + 80 T_F \Nf
\right) \frac{a_{\mbox{\footnotesize{$\MSbars$}}}}{36} \right. \nonumber \\
&& \left. ~+~ \left( \left( 18 \alpha_{\mbox{\footnotesize{$\MSbars$}}}^4
- 18 \alpha_{\mbox{\footnotesize{$\MSbars$}}}^3
+ 190 \alpha_{\mbox{\footnotesize{$\MSbars$}}}^2
- 576 \zeta(3) \alpha_{\mbox{\footnotesize{$\MSbars$}}}
+ 463 \alpha_{\mbox{\footnotesize{$\MSbars$}}} + 864 \zeta(3) - 7143 \right)
C_A^2 \right. \right. \nonumber \\
&& \left. \left. ~~~~~~~+~ \left( -~ 320
\alpha_{\mbox{\footnotesize{$\MSbars$}}}^2
- 320 \alpha_{\mbox{\footnotesize{$\MSbars$}}} + 2304 \zeta(3)
+ 4248 \right) C_A T_F \Nf \right. \right. \nonumber \\
&& \left. \left. ~~~~~~~+~ \left( -~ 4608 \zeta(3) + 5280 \right) C_F T_F \Nf
\right) \frac{a^2_{\mbox{\footnotesize{$\MSbars$}}}}{288} \right]
\alpha_{\mbox{\footnotesize{$\MSbars$}}} ~+~ O \left(
a^3_{\mbox{\footnotesize{$\MSbars$}}} \right) ~. 
\end{eqnarray}
These relations are important as we will be providing results in arbitrary
linear covariant gauge in each scheme separately. Therefore the conversion of
parameters between schemes has to be taken into account when comparing results
and, moreover, when constructing conversion functions for the operators. 

One reason we provide the amplitudes in both the $\MSbar$ and 
RI${}^\prime$/SMOM schemes is that it turns out that there is no definitive 
scheme definition for the latter. This is because in discussing the general 
scheme definition we alluded to channel $0$ which represents some combination 
of the scalar amplitudes. This choice is not unique as indicated in \cite{31}. 
For instance, one could extract the combination given by multiplying the 
Green's function by the Born term which was the approach of \cite{27,29} for 
the tensor current. However, one could equally ensure that that amplitude with 
the poles in $\epsilon$ has no $O(a)$ terms after renormalization. Indeed this 
statement does depend on how one defined the basis tensors and therefore one 
has a range of choices as to how to define RI${}^\prime$/SMOM. Though as an 
aside the renormalization of $\partial W_2$ has to be carried out in such a way
that it is consistent with the Slavnov-Taylor identity, \cite{31}, since this 
operator is related to the divergence of the vector current which is conserved 
and a physical operator with zero anomalous dimension to all orders in 
perturbation theory in all renormalization schemes. The version of the 
RI${}^\prime$/SMOM scheme which we will use here is a direct extension of the 
one loop version given in \cite{31}. There the coefficients of the channels $1$
and $2$ were used to define the $W_2$ renormalization constant. Two channels 
were used due to the fact that we have two renormalization constants to fix for
the first row of the mixing matrix of (\ref{mixmat}). As with the three loop 
$\MSbar$ renormalization of \cite{32} the counterterms are entwined with each
other and so one has to solve a set of linear equations to fix the explicit 
values for each renormalization constant. Therefore, all the amplitudes which 
we record for RI${}^\prime$/SMOM are determined via that condition. Though we 
emphasise that this choice is not unique in order to define the renormalization
constants. One could have, for instance, multiplied the Green's function by the
Born term. Alternatively one could use the same combination of amplitudes as
was used for $\partial W_2$, \cite{31}, for consistency with the Slavnov-Taylor
identity of that operator. Though that would not be sufficient on its own as
one would require a second independent projection to solve for both elements of 
(\ref{mixmat}). However, in order to assist with lattice computations in the
situation where a different amplitude combination might be made to renormalize 
the operator, we provide the $\MSbar$ amplitudes as it is the canonical 
reference scheme as well as to facilitate making different variations on the
scheme definition. This is because one can then readily convert from any scheme
to $\MSbar$ for comparison. Indeed some lattice groups carry out their 
renormalization according to their own prescription before converting their 
results to $\MSbar$ prior to performing the actual matching to the continuum 
results.

Finally, having described the theoretical background to the problem we note
how it was implemented in practical terms. The main tool used to handle the
algebra of Lorentz indices, projection matrices and evaluation of the 
underlying Feynman graphs was the symbolic manipulation language {\sc Form}, 
\cite{41}. The actual Feynman diagrams themselves were constructed in 
electronic format by using the {\sc Qgraf} package, \cite{42}. These were then 
converted into the notation used for the {\sc Form} computations where the 
colour and Lorentz indices were added. At one loop there were $3$ graphs and at
two loops there were $37$ diagrams. The algorithm to evaluate the integrals 
comprising each Feynman graph was to first write them as scalar integrals. By 
this we mean in a form which is the starting point of the Laporta algorithm. 
Each integral is rewritten in terms of the denominator propagators as far as 
possible. For those cases where this is not possible the numerator tensor 
structure is written in terms of irreducible propagators. In this form one can 
apply the Laporta method, \cite{43}, where a redundant set of linear equations 
is established between all the integrals by using integration by parts and 
Lorentz identities. These relations are then solved to express each required 
integral in terms of a small set of scalar master integrals. These have been 
evaluated directly and given in a set of articles, \cite{44,45,46,47}. A 
summary has been provided in \cite{29}. Central to the construction of the 
Laporta algorithm used for the current problem was the use of the {\sc Reduze} 
package, \cite{48}, which is based on the {\sc Ginac} system, \cite{49}, which 
is written in C$++$. For the two loop Feynman integrals we needed to evaluate, 
it turns out that there are only two basic momentum topologies for which one 
needs to construct the reduction of integrals. These are the ladder and its 
associated non-planar ladder. The momentum routing of all the Feynman graphs 
could be mapped to these two topologies and therefore, aside from the 
elementary one loop case, {\sc Reduze} was only required to build two sets of
integration by parts and Lorentz identity relations. The machinery derived for 
this set of operators was also used in constructing the two loop amplitudes for
the quark currents, \cite{30}. There the anomalous dimensions for the scalar 
and tensor currents reproduced the known results of \cite{27,28,29} and 
therefore this provides us with a strong check on the routines which have been 
built and used for level $W_2$ here. Finally, in order to carry out the 
renormalization prior to deducing the values of the amplitudes, we follow the
method of \cite{50} for automatic calculations. We perform the computation for
bare parameters and operators and then introduce the counterterms by rescaling
all bare quantities with their renormalized equivalences. For the operators
themselves this ethos also applies in that the bare operators are replaced by
their renormalized counterparts defined by the structure of the mixing matrix,
(\ref{mixmat}).  Then the counterterms for the operators are fixed with respect
to whichever is the scheme prescription after the known counterterms for the 
coupling constant and gauge parameter have been included from previous 
calculations.  

\sect{Amplitudes.}

This section is devoted to recording the amplitudes for the two second moment 
flavour non-singlet Wilson operators, $W_2$ and $\partial W_2$, inserted in the
quark $2$-point function. As there are quite a large number of amplitudes and 
two operators to provide the full set of expressions for in each scheme, we
present the results of the finite parts in a set of Tables. We follow the same 
notation as \cite{30} in recording the sets of numbers which appear as 
coefficients of both the group Casimirs as well as a set of basis numbers for
each amplitude. The latter derive from the explicit values of the basic two 
loop scalar master Feynman diagrams which were summarized in \cite{29}. More 
concretely our amplitudes are written in general as 
\begin{eqnarray}
\Sigma^{{\cal O}^i}_{(i)}(p,q) &=& \left( \sum_n c^{{\cal O}^i,(1)}_{(i)\,n} 
a^{(1)}_n \right) C_F a ~+~ 
\left( \sum_n c^{{\cal O}^i,(21)}_{(i)\,n} a^{(21)}_n \right) C_F T_F \Nf a^2 
\nonumber \\
&& +~ \left( \sum_n c^{{\cal O}^i,(22)}_{(i)\,n} a^{(22)}_n \right) 
C_F C_A a^2 ~+~ \left( \sum_n c^{{\cal O}^i,(23)}_{(i)\,n} a^{(23)}_n \right) 
C_F^2 a^2 ~+~ O(a^3) ~.
\end{eqnarray}
Here the $a^{(l)}_n$ are the basis of numbers which for simplicity also
includes the gauge parameter. The label $l$ here indicates both the loop order,
as the first number in the two loop case, whilst the second number at two loops
relates to a specific colour group Casimir. The coefficients 
$c^{{\cal O}^i,(l)}_{(i)\,n}$ are the actual rational numbers which appear in 
the appropriate piece of the amplitude as is evident from the Tables 
themselves. Clearly these reduce the space needed to display the explicit 
results and allows for easier comparison of the structure of the amplitudes 
between schemes. For the numbers in the basis we use the notation of \cite{29}.
For instance, $\psi(z)$ is the derivative of the logarithm of the Euler Gamma 
function and 
\begin{equation}
s_n(z) ~=~ \frac{1}{\sqrt{3}} \Im \left[ \mbox{Li}_n \left( 
\frac{e^{iz}}{\sqrt{3}} \right) \right]
\end{equation}
where $\mbox{Li}_n(z)$ is the polylogarithm function. The quantity $\Sigma$ is 
a combination of various harmonic polylogarithms, \cite{29,47},
\begin{equation}
\Sigma ~=~ {\cal H}^{(2)}_{31} ~+~ {\cal H}^{(2)}_{43} 
\end{equation}
and this combination always appears.

Whilst the Tables\footnote{Attached to this article is an electronic file where
all the expressions presented in the Tables are available in a useable format.}
reflect the size of the computation undertaken for more practical purposes it 
is appropriate to provide the explicit numerical evaluation of the results. In 
order to achieve this we note that the explicit numerical values of the various
numbers in the basis are taken to be 
\begin{eqnarray}
\zeta(3) &=& 1.20205690 ~~,~~ \Sigma ~=~ 6.34517334 ~~,~~ 
\psi^\prime\left( \frac{1}{3} \right) ~=~ 10.09559713 \nonumber \\
\psi^{\prime\prime\prime}\left( \frac{1}{3} \right) &=& 488.1838167 ~~,~~ 
s_2\left( \frac{\pi}{2} \right) ~=~ 0.32225882 ~~,~~ 
s_2\left( \frac{\pi}{6} \right) ~=~ 0.22459602 \nonumber \\
s_3\left( \frac{\pi}{2} \right) &=& 0.32948320 ~~,~~ 
s_3\left( \frac{\pi}{6} \right) ~=~ 0.19259341 ~.
\end{eqnarray}
Therefore, we can record the numerical values in both schemes. We have chosen 
to do this for the case of the $SU(3)$ colour group. For the $\MSbar$ scheme we
have 
\begin{eqnarray}
\left. \Sigma^{W_2}_{(1)}(p,q) \right|_{\MSbars} &=& -~ \left[ 1.7499534 
+ 0.4444444 \alpha \right] a \nonumber \\
&& -~ \left[ 37.3849283 + 2.4296818 \alpha + 1.6608855 \alpha^2 - 5.0243682 \Nf
\right] a^2 ~+~ O(a^3) \nonumber \\
\left. \Sigma^{W_2}_{(2)}(p,q) \right|_{\MSbars} &=& -~ 1 ~+~ \left[ 
3.3748835 - 0.1387491 \alpha \right] a \nonumber \\
&& +~ \left[ 43.5097605 - 0.7932193 \alpha - 0.1511137 \alpha^2 
- 4.7881095 \Nf \right] a^2 ~+~ O(a^3) \nonumber \\
\left. \Sigma^{W_2}_{(3)}(p,q) \right|_{\MSbars} &=& \left[ 1.2531175 
+ 0.2037969 \alpha \right] a \nonumber \\
&& +~ \left[ 15.7237696 + 3.9484943 \alpha + 0.7217806 \alpha^2 
- 0.9599542 \Nf \right] a^2 ~+~ O(a^3) \nonumber \\
\left. \Sigma^{W_2}_{(4)}(p,q) \right|_{\MSbars} &=& \left[ 1.6541967
+ 0.4075938 \alpha \right] a \nonumber \\
&& +~ \left[ 16.8558938 + 4.8339947 \alpha + 1.4319988 \alpha^2 
- 1.7451297 \Nf \right] a^2 ~+~ O(a^3) \nonumber \\
\left. \Sigma^{W_2}_{(5)}(p,q) \right|_{\MSbars} &=& \left[ 2.6233015 
+ 1.7040764 \alpha \right] a \nonumber \\
&& +~ \left[ 60.6554860 + 15.1801309 \alpha + 6.1487315 \alpha^2 
- 4.8269741 \Nf \right] a^2 ~+~ O(a^3) \nonumber \\
\left. \Sigma^{W_2}_{(6)}(p,q) \right|_{\MSbars} &=& \left[ 0.9322541 
+ 0.6850920 \alpha \right] a \nonumber \\
&& +~ \left[ 28.1056849 + 9.0379699 \alpha + 2.4124553 \alpha^2 
- 0.9014209 \Nf \right] a^2 ~+~ O(a^3) \nonumber \\
\left. \Sigma^{W_2}_{(7)}(p,q) \right|_{\MSbars} &=& \left[ 1.5956635 
+ 1.0926858 \alpha \right] a \nonumber \\
&& +~ \left[ 46.3221825 + 11.8708374 \alpha + 3.9440029 \alpha^2 
- 2.4490197 \Nf \right] a^2 ~+~ O(a^3) \nonumber \\
\left. \Sigma^{W_2}_{(8)}(p,q) \right|_{\MSbars} &=& \left[ 1.6910474 
+ 0.4075938 \alpha \right] a \nonumber \\
&& +~ \left[ 21.8712121 + 5.2430694 \alpha + 1.4690359 \alpha^2 
- 1.6999495 \Nf \right] a^2 ~+~ O(a^3) \nonumber \\
\left. \Sigma^{W_2}_{(9)}(p,q) \right|_{\MSbars} &=& -~ 0.4444444 a 
\nonumber \\
&& -~ \left[ 8.8385249 + 1.3267080 \alpha + 0.0370370 \alpha^2 
- 0.4428779 \Nf \right] a^2 ~+~ O(a^3) \nonumber \\
\left. \Sigma^{W_2}_{(10)}(p,q) \right|_{\MSbars} &=& -~ 1.6390287 a 
\nonumber \\
&& -~ \left[ 30.9488446 - 0.9782418 \alpha + 0.1365857 \alpha^2 
- 2.5665832 \Nf \right] a^2 \nonumber \\
&& +~ O(a^3) ~. 
\end{eqnarray}
Those for $\partial W_2$ are 
\begin{eqnarray}
\left. \Sigma^{\partial W_2}_{(1)}(p,q) \right|_{\MSbars} &=& 
\left. \Sigma^{\partial W_2}_{(2)}(p,q) \right|_{\MSbars} \nonumber \\
&=& -~ 1 ~+~ \left[ 1.6249301 - 0.5831936 \alpha \right] a \nonumber \\
&& +~ \left[ 6.1248321 - 3.2229010 \alpha - 1.8119992 \alpha^2 
+ 0.2362586 \Nf \right] a^2 ~+~ O(a^3) \nonumber \\
\left. \Sigma^{\partial W_2}_{(3)}(p,q) \right|_{\MSbars} &=& 
\left. \Sigma^{\partial W_2}_{(8)}(p,q) \right|_{\MSbars} \nonumber \\
&=& \left[ 2.9441649 + 0.6113907 \alpha \right] a \nonumber \\
&& +~ \left[ 37.5949817 + 9.1915636 \alpha + 2.1908165 \alpha^2 
- 2.6599037 \Nf \right] a^2 ~+~ O(a^3) \nonumber \\
\left. \Sigma^{\partial W_2}_{(4)}(p,q) \right|_{\MSbars} &=& 
\left. \Sigma^{\partial W_2}_{(7)}(p,q) \right|_{\MSbars} \nonumber \\
&=& \left[ 3.2498602 + 1.5002795 \alpha \right] a \nonumber \\
&& +~ \left[ 63.1780763 + 16.7048322 \alpha + 5.3760017 \alpha^2 
- 4.1941494 \Nf \right] a^2 ~+~ O(a^3) \nonumber \\ 
\left. \Sigma^{\partial W_2}_{(5)}(p,q) \right|_{\MSbars} &=& 
\left. \Sigma^{\partial W_2}_{(6)}(p,q) \right|_{\MSbars} \nonumber \\
&=& \left[ 3.5555556 + 2.3891684 \alpha \right] a \nonumber \\
&& +~ \left[ 88.7611709 + 24.2181008 \alpha + 8.5611869 \alpha^2 
- 5.7283951 \Nf \right] a^2 ~+~ O(a^3) \nonumber \\ 
\left. \Sigma^{\partial W_2}_{(9)}(p,q) \right|_{\MSbars} &=& 
\left. \Sigma^{\partial W_2}_{(10)}(p,q) \right|_{\MSbars} \nonumber \\
&=& -~ 2.0834731 a \nonumber \\
&& -~ \left[ 39.7873696 + 0.3484662 \alpha + 0.1736228 \alpha^2 
- 3.0094611 \Nf \right] a^2 \nonumber \\
&& +~ O(a^3) ~.
\end{eqnarray}
The explicit results from which these numerical values were derived are 
recorded in Tables $1$ to $12$. However, the equivalence between amplitudes
indicated above for $\partial W_2$ are exact at two loops which is why we have
not included parallel columns with the same values. Indeed these equivalences
are a minor check on our computation as the tensor basis is clearly reflection
symmetric when the operator insertion has the same property as is the case for
$\partial W_2$. Another check resides in the fact that some of the amplitudes
for $\partial W_2$ are proportional to the two loop amplitudes for the vector 
current of \cite{30}. For instance,
\begin{eqnarray}
\left. \Sigma^V_{(1)}(p,q) \right|_{\MSbars} &=& 
\left. \Sigma^{\partial W_2}_{(1)}(p,q) \right|_{\MSbars} ~~,~~  
\left. \Sigma^V_{(2)}(p,q) \right|_{\MSbars} ~=~ \half 
\left. \Sigma^{\partial W_2}_{(3)}(p,q) \right|_{\MSbars} \nonumber \\
\left. \Sigma^V_{(3)}(p,q) \right|_{\MSbars} &=& \half 
\left. \Sigma^{\partial W_2}_{(5)}(p,q) \right|_{\MSbars} ~~,~~  
\left. \Sigma^V_{(6)}(p,q) \right|_{\MSbars} ~=~ \half 
\left. \Sigma^{\partial W_2}_{(9)}(p,q) \right|_{\MSbars} 
\label{veceqms}
\end{eqnarray} 
where $V$ indicates the vector current in the notation of \cite{31,32}. Though 
it should be stressed that the associated amplitude basis tensors of each of 
the operators are not the same. This is trivial to see because of the mismatch 
in the number of Lorentz indices on each of the operators. Indeed this is why 
channel $4$ of $\partial W_2$ does not feature in (\ref{veceqms}) as this index
imbalance means that there are not the same number of amplitudes for each
operator. 

To illustrate that the Slavnov-Taylor identity is satisfied by the operator 
$\partial W_2$ we have to project out that part of the Green's function which 
contains the divergence of the vector operator and was discussed in 
\cite{30,31}. This is achieved by simply contracting the two free indices of 
the $\partial W_2$ operator. From the explicit forms of the amplitudes in the 
Tables the combination proportional to $\pslash$ is
\begin{eqnarray}
&& -~ \frac{[d-2]}{d} \left. \Sigma^{\partial W_2}_{(1)}(p,q)
\right|_{\MSbars} ~-~
\left. \Sigma^{\partial W_2}_{(2)}(p,q) \right|_{\MSbars} ~+~
\frac{[d-4]}{4d} \left. \Sigma^{\partial W_2}_{(3)}(p,q) \right|_{\MSbars}
\nonumber \\
&& +~ \frac{[d+2]}{2d} \left. \Sigma^{\partial W_2}_{(4)}(p,q)
\right|_{\MSbars} ~+~
\frac{[d-4]}{4d} \left. \Sigma^{\partial W_2}_{(5)}(p,q) \right|_{\MSbars}
\nonumber \\
&& =~ \frac{3}{2} ~+~ \frac{3}{2} C_F \alpha a \nonumber \\
&& ~~~~+~ C_F \left[ \left[ \frac{123}{8} + \frac{39}{4} \alpha + \frac{27}{16}
\alpha^2 - \frac{9}{2} \zeta(3) - \frac{9}{2} \zeta(3) \alpha \right] C_A
\right. \nonumber \\
&& \left. ~~~~~~~~~~~~~~-~ \frac{21}{4} T_F \NF ~-~ \frac{15}{16} C_F 
\right] a^2 ~+~ O(a^3) ~.
\label{stidms}
\end{eqnarray}
Clearly the right hand side is proportional to the finite part of the quark 
$2$-point functions after renormalization in the $\MSbar$ scheme. A similar
combination gives the part proportional to $\qslash$ with the same expression 
to two loops. We emphasise that the $\MSbar$ renormalization constant used to
renormalize $\partial W_2$ is the same as was used in \cite{32} to construct
the operator correlation functions.  
 
For the RI${}^\prime$/SMOM scheme we have to be careful to define the
renormalization of $\partial W_2$ so that the Slavnov-Taylor identity is
satisfied. In other words the renormalization constant for $\partial W_2$ is
already determined since the vector current is a physical operator and 
therefore its renormalization constant is unity in {\em all} schemes. However, 
for $W_2$ itself one has a large degree of freedom to define the 
RI${}^\prime$/SMOM scheme renormalization constant. This was discussed in 
\cite{30,31} and we follow the prescription used there. This was to fix the 
finite parts of the other renormalization constants of the $W_2$ level mixing 
matrix so that for those channels containing the poles in $\epsilon$ there was 
no $O(a)$ corrections. Of course this is not the unique way of defining these 
renormalization constants. One could, for example, take some sort of projection
of the Green's function and require that there is no $O(a)$ correction for that
particular combination. The fact that there is a degree of ambiguity is one of 
the reasons why we have provided the $\MSbar$ results. Given that we are 
extending \cite{31} to two loops we follow that prescription here and record 
those results. The complete results are in Tables $13$ to $23$ and the $SU(3)$ 
numerical results for the operator $W_2$ itself are
\begin{eqnarray}
\Sigma^{W_2}_{(3)}(p,q) &=& \left[ 1.2531175 + 0.2037969 \alpha \right] a 
\nonumber \\
&& +~ \left[ 16.9936409 + 4.6449275 \alpha + 0.8180047 \alpha^2 
+ 0.1528477 \alpha^3 \right. \nonumber \\
&& \left. ~~~~
-~ ( 0.9599542 + 0.2264410 \alpha )\Nf \right] a^2 ~+~ O(a^3) \nonumber \\
\Sigma^{W_2}_{(4)}(p,q) &=& \left[ 1.6541967 + 0.4075938 \alpha \right] a 
\nonumber \\
&& +~ \left[ 19.6462784 + 6.6534412 \alpha + 1.5011981 \alpha^2 
+ 0.3056953 \alpha^3 \right. \nonumber \\
&& \left. ~~~~
-~ ( 1.7451297 + 0.4528820 \alpha )\Nf \right] a^2 ~+~ O(a^3) \nonumber \\
\Sigma^{W_2}_{(5)}(p,q) &=& \left[ 2.6233015 + 1.7040764 \alpha \right] a 
\nonumber \\
&& +~ \left[ 67.8774218 + 32.7286131 \alpha + 8.1639217 \alpha^2 
+ 1.2780573 \alpha^3 \right. \nonumber \\
&& \left. ~~~~
-~ ( 4.8269741 + 1.8934182 \alpha )\Nf \right] a^2 ~+~ O(a^3) \nonumber \\
\Sigma^{W_2}_{(6)}(p,q) &=& \left[ 0.9322541 + 0.6850920 \alpha \right] a 
\nonumber \\
&& +~ \left[ 26.6612784 + 12.6105871 \alpha + 2.5876701 \alpha^2 
+ 0.5138190 \alpha^3 \right. \nonumber \\
&& \left. ~~~~
-~ ( 0.9014209 + 0.7612133 \alpha )\Nf \right] a^2 ~+~ O(a^3) \nonumber \\
\Sigma^{W_2}_{(7)}(p,q) &=& \left[ 1.5956635 + 1.0926858 \alpha \right] a 
\nonumber \\
&& +~ \left[ 48.8125936 + 22.7212025 \alpha + 5.2502695 \alpha^2 
+ 0.8195143 \alpha^3 \right. \nonumber \\
&& \left. ~~~~
-~ ( 2.4490197 + 1.2140953 \alpha )\Nf \right] a^2 ~+~ O(a^3) \nonumber \\
\Sigma^{W_2}_{(8)}(p,q) &=& \left[ 1.6910474 + 0.4075938 \alpha \right] a 
\nonumber \\
&& +~ \left[ 25.3854030 + 8.7651597 \alpha + 1.9332969 \alpha^2 
+ 0.3056953 \alpha^3 \right. \nonumber \\
&& \left. ~~~~
-~ ( 1.6999495 + 0.4528820 \alpha )\Nf \right] a^2 ~+~ O(a^3) \nonumber \\
\Sigma^{W_2}_{(9)}(p,q) &=& -~ 0.4444444 a \nonumber \\
&& -~ \left[ 7.4702494 + 0.5365824 \alpha + 0.0370370 \alpha^2 
- 0.4428779 \Nf \right] a^2 ~+~ O(a^3) \nonumber \\
\Sigma^{W_2}_{(10)}(p,q) &=& -~ 1.6390287 a \nonumber \\
&& -~ \left[ 35.7026184 - 1.4031864 \alpha + 0.1365857 \alpha^2 
- 2.5665832 \Nf \right] a^2 \nonumber \\
&& +~ O(a^3) ~. 
\end{eqnarray}
Those for the total derivative operator are 
\begin{eqnarray}
\Sigma^{\partial W_2}_{(1)}(p,q) &=& \Sigma^{\partial W_2}_{(2)}(p,q) 
\nonumber \\
&=& -~ 1 ~+~ \left[ 1.6249301 + 0.7501398 \alpha \right] a \nonumber \\
&& +~ \left[ 31.5890382 + 12.2494724 \alpha + 2.8130241 \alpha^2 
+ 0.5626048 \alpha^3 \right. \nonumber \\
&& \left. ~~~~
-~ ( 2.0970747 + 0.8334886 \alpha )\Nf \right] a^2 ~+~ O(a^3) \nonumber \\
\Sigma^{\partial W_2}_{(3)}(p,q) &=& 
\Sigma^{\partial W_2}_{(8)}(p,q) \nonumber \\
&=& \left[ 2.9441649 + 0.6113907 \alpha \right] a \nonumber \\
&& +~ \left[ 37.5949817 + 10.2080849 \alpha + 2.2927149 \alpha^2 
+ 0.4585430 \alpha^3 \right. \nonumber \\
&& \left. ~~~~
-~ ( 2.6599037 + 0.6793229 \alpha )\Nf \right] a^2 ~+~ O(a^3) \nonumber \\
\Sigma^{\partial W_2}_{(4)}(p,q) &=& 
\Sigma^{\partial W_2}_{(7)}(p,q) \nonumber \\
&=& \left[ 3.2498602 + 1.5002795 \alpha \right] a \nonumber \\
&& +~ \left[ 63.1780763 + 24.4989449 \alpha + 5.6260483 \alpha^2 
+ 1.1252097 \alpha^3 \right. \nonumber \\
&& \left. ~~~~
-~ ( 4.1941494 + 1.6669773 \alpha )\Nf \right] a^2 ~+~ O(a^3) \nonumber \\
\Sigma^{\partial W_2}_{(5)}(p,q) &=& 
\Sigma^{\partial W_2}_{(6)}(p,q) \nonumber \\
&=& \left[ 3.5555556 + 2.3891684 \alpha \right] a \nonumber \\
&& +~ \left[ 88.7611709 + 38.7898049 \alpha + 8.9593816 \alpha^2 
+ 1.7918763 \alpha^3 \right. \nonumber \\
&& \left. ~~~~
-~ ( 5.7283951 + 2.6546316 \alpha )\Nf \right] a^2 ~+~ O(a^3) \nonumber \\
\Sigma^{\partial W_2}_{(9)}(p,q) &=& 
\Sigma^{\partial W_2}_{(10)}(p,q) \nonumber \\
&=& -~ 2.0834731 a \nonumber \\
&& -~ \left[ 39.7873696 - 2.4294979 \alpha + 0.1736228 \alpha^2 
- 3.0094611 \Nf \right] a^2 \nonumber \\
&& +~ O(a^3) ~.
\end{eqnarray}
The same remarks made for the $\MSbar$ case concerning the equivalences 
indicated above between amplitudes for $\partial W_2$ and their proportionality
with those of the vector operator of \cite{30} also apply to the 
RI${}^\prime$/SMOM scheme results. So the relations of (\ref{veceqms}) are true
when the $\MSbar$ indication is dropped. Equally we note that the 
Slavnov-Taylor identity is also satisfied in the RI${}^\prime$/SMOM scheme. The
parallel computation to
(\ref{stidms}) is 
\begin{eqnarray}
&& -~ \frac{[d-2]}{d} \Sigma^{\partial W_2}_{(1)}(p,q) ~-~
\Sigma^{\partial W_2}_{(2)}(p,q) ~+~
\frac{[d-4]}{4d} \Sigma^{\partial W_2}_{(3)}(p,q) \nonumber \\
&& +~ \frac{[d+2]}{2d} \Sigma^{\partial W_2}_{(4)}(p,q) ~+~
\frac{[d-4]}{4d} \Sigma^{\partial W_2}_{(5)}(p,q) ~=~ \frac{3}{2} ~+~ O(a^3) 
\label{stidsmom}
\end{eqnarray}
to the order we have computed to. As the quark wave function renormalization
constant is in the RI${}^\prime$ scheme there are no $O(a)$ corrections which
indicates consistency with the Slavnov-Taylor identity. As a final observation
on the two loop values of the amplitudes we note that the relation noted in 
\cite{31} at one loop
\begin{equation}
\Sigma^{\partial W_2}_{(3)}(p,q) ~-~ 2 \Sigma^{\partial W_2}_{(4)}(p,q) ~+~
\Sigma^{\partial W_2}_{(5)}(p,q) ~=~ O(a^3)
\end{equation}
is valid to two loops for all $\alpha$ and for both the $\MSbar$ and 
RI${}^\prime$/SMOM schemes.

\sect{Anomalous dimensions.}

In order to record these amplitudes we have determined the renormalization 
constants which were fixed by the RI${}^\prime$/SMOM scheme choice defined in
\cite{31}. These are encoded in the mixing matrix of anomalous dimensions which
extends (\ref{anomdimms}) and is
\begin{eqnarray}
\left. \frac{}{} \gamma^{W_2}_{11}(a,\alpha) 
\right|_{\mbox{\footnotesize{RI$^\prime$/SMOM}}} &=& \frac{8}{3} C_F a 
\nonumber \\
&& +~ \left[ \left[ ( 108 \alpha^2 + 324 \alpha - 924 ) \psi^\prime(\third)
- ( 72 \alpha^2 + 216 \alpha - 616 ) \pi^2 \right. \right. \nonumber \\
&& \left. \left. ~~~~~-~ 81 \alpha^2 - 243 \alpha + 16866 \right] C_A ~-~
2016 C_F \right. \nonumber \\
&& \left. ~~~~+ \left[ 336 \psi^\prime(\third) - 224 \pi^2 - 5976 \right]
T_F \Nf \right] \frac{C_F a^2}{486} ~+~ O(a^3) \nonumber \\
\left. \frac{}{} \gamma^{W_2}_{12}(a,\alpha) 
\right|_{\mbox{\footnotesize{RI$^\prime$/SMOM}}} &=& -~ \frac{4}{3} C_F a 
\nonumber \\
&& +~ \left[ \left[ 264 \psi^\prime(\third) - 176 \pi^2 - 81 \alpha^2 
- 243 \alpha - 6651 \right] C_A \right. \nonumber \\
&& \left. ~~~~+ \left[ 144 \psi^\prime(\third) - 288 \psi^\prime(\third)
\alpha - 288 + 648 \alpha - 96 \pi^2 + 192 \pi^2 \alpha^2 \right] C_F \right.
\nonumber \\
&& \left. ~~~~+ \left[ 64 \pi^2 + 2340 - 96 \psi^\prime(\third) \right]
T_F \Nf \right] \frac{C_F a^2}{486} ~+~ O(a^3) \nonumber \\
\left. \frac{}{} \gamma^{W_2}_{22}(a,\alpha) 
\right|_{\mbox{\footnotesize{RI$^\prime$/SMOM}}} &=&  O(a^3) ~. 
\end{eqnarray}
As a check on the derivation of these we ensured that the two loop $\MSbar$
matrix of anomalous dimensions emerged from the same {\sc Form} programme. They
agreed with the original computation of \cite{32} which provided the matrix at
three loops. However, that computation was performed with the use of the 
{\sc Form} version of the {\sc Mincer} algorithm, \cite{51,52}. There one could
deduce the off-diagonal matrix element by choosing to route the single external
momentum out through the operator insertion itself in contrast to the current 
computation where there are two independent momenta. As $\partial W_2$ is in 
essence the vector operator then the $22$ element is equivalent to the vector 
operator anomalous dimension. 

The RI${}^\prime$/SMOM scheme anomalous dimensions can be computed by a second 
method involving conversion functions. For an introduction to these see 
\cite{35}. However, for level $W_2$ this is not as straightforward since one is
not dealing with a multiplicatively renormalizable operator. Instead there is a
matrix of renormalization constants and therefore the concept of a conversion 
function translates into a {\em matrix} of conversion functions. Formally, for 
$W_2$ this is 
\begin{equation}
C^{W_2}_{ij}(a,\alpha) ~=~ 
Z^{W_2}_{{ik},\mbox{\footnotesize{RI$^\prime$/SMOM}}}
\left[ Z^{W_2}_{{kj},\mbox{\footnotesize{$\MSbar$}}} \right]^{-1} ~.
\end{equation} 
The explicit forms of the elements of this matrix are straightforward to deduce
and are 
\begin{eqnarray}
C^{W_2}_{11}(a,\alpha) &=&
\frac{Z^{W_2}_{11,\mbox{\footnotesize{RI$^\prime$/SMOM}}}}
{Z^{W_2}_{11,\mbox{\footnotesize{$\MSbar$}}}} ~~~,~~~ 
C^{W_2}_{22}(a,\alpha) ~=~
\frac{Z^{W_2}_{22,\mbox{\footnotesize{RI$^\prime$/SMOM}}}}
{Z^{W_2}_{22,\mbox{\footnotesize{$\MSbar$}}}} \nonumber \\
C^{W_2}_{12}(a,\alpha) &=&
\frac{Z^{W_2}_{12,\mbox{\footnotesize{RI$^\prime$/SMOM}}}}
{Z^{W_2}_{22,\mbox{\footnotesize{$\MSbar$}}}} ~-~ 
\frac{Z^{W_2}_{11,\mbox{\footnotesize{RI$^\prime$/SMOM}}}
Z^{W_2}_{12,\mbox{\footnotesize{$\MSbar$}}}} 
{Z^{W_2}_{11,\mbox{\footnotesize{$\MSbar$}}}
Z^{W_2}_{22,\mbox{\footnotesize{$\MSbar$}}}} 
\end{eqnarray}
where $C^{W_2}_{21}(a,\alpha)$~$=$~$0$ from the upper triangular nature of the
underlying renormalization matrices. Indeed this means that the diagonal
conversion functions are what one might naively expect. From this one can 
derive the relation between the anomalous dimensions in both schemes.
Specifically we have 
\begin{eqnarray}
\gamma^{W_2}_{11,\mbox{\footnotesize{RI$^\prime$/SMOM}}}
\left(a_{\mbox{\footnotesize{RI$^\prime$}}},
\alpha_{\mbox{\footnotesize{RI$^\prime$}}}\right) &=&
\gamma^{W_2}_{11,\mbox{\footnotesize{$\MSbar$}}}
\left(a_{\mbox{\footnotesize{$\MSbar$}}}\right) ~-~
\beta\left(a_{\mbox{\footnotesize{$\MSbar$}}}\right)
\frac{\partial ~}{\partial a_{\mbox{\footnotesize{$\MSbar$}}}}
\ln C^{W_2}_{11}\left(a_{\mbox{\footnotesize{$\MSbar$}}},
\alpha_{\mbox{\footnotesize{$\MSbar$}}}\right) \nonumber \\
&& -~ \alpha_{\mbox{\footnotesize{$\MSbar$}}}
\gamma^{\mbox{\footnotesize{$\MSbar$}}}_\alpha
\left(a_{\mbox{\footnotesize{$\MSbar$}}},
\alpha_{\mbox{\footnotesize{$\MSbar$}}}\right)
\frac{\partial ~}{\partial \alpha_{\mbox{\footnotesize{$\MSbar$}}}}
\ln C^{W_2}_{11}\left(a_{\mbox{\footnotesize{$\MSbar$}}},
\alpha_{\mbox{\footnotesize{$\MSbar$}}}\right)
\end{eqnarray}
\begin{eqnarray}
\gamma^{W_2}_{12,\mbox{\footnotesize{RI$^\prime$/SMOM}}}
\left(a_{\mbox{\footnotesize{RI$^\prime$}}},
\alpha_{\mbox{\footnotesize{RI$^\prime$}}}\right) &=&
\left[ \gamma^{W_2}_{12,\mbox{\footnotesize{$\MSbar$}}}
\left(a_{\mbox{\footnotesize{$\MSbar$}}}\right)
C^{W_2}_{11}\left(a_{\mbox{\footnotesize{$\MSbar$}}},
\alpha_{\mbox{\footnotesize{$\MSbar$}}}\right) \right. \nonumber \\
&& \left. ~-
\beta\left(a_{\mbox{\footnotesize{$\MSbar$}}}\right)
\frac{\partial ~}{\partial a_{\mbox{\footnotesize{$\MSbar$}}}}
C^{W_2}_{12}\left(a_{\mbox{\footnotesize{$\MSbar$}}},
\alpha_{\mbox{\footnotesize{$\MSbar$}}}\right) \right. \nonumber \\
&& \left. ~- \alpha_{\mbox{\footnotesize{$\MSbar$}}}
\gamma^{\mbox{\footnotesize{$\MSbar$}}}_\alpha
\left(a_{\mbox{\footnotesize{$\MSbar$}}},
\alpha_{\mbox{\footnotesize{$\MSbar$}}}\right)
\frac{\partial ~}{\partial \alpha_{\mbox{\footnotesize{$\MSbar$}}}}
C^{W_2}_{12}\left(a_{\mbox{\footnotesize{$\MSbar$}}},
\alpha_{\mbox{\footnotesize{$\MSbar$}}}\right) \right. \nonumber \\
&& \left. ~- \gamma^{W_2}_{11,\mbox{\footnotesize{$\MSbar$}}}
\left(a_{\mbox{\footnotesize{$\MSbar$}}}\right)
C^{W_2}_{12}\left(a_{\mbox{\footnotesize{$\MSbar$}}},
\alpha_{\mbox{\footnotesize{$\MSbar$}}}\right) \right. \nonumber \\
&& \left. ~+ \gamma^{W_2}_{22,\mbox{\footnotesize{$\MSbar$}}}
\left(a_{\mbox{\footnotesize{$\MSbar$}}}\right)
C^{W_2}_{12}\left(a_{\mbox{\footnotesize{$\MSbar$}}},
\alpha_{\mbox{\footnotesize{$\MSbar$}}}\right) \right. \nonumber \\
&& \left. ~+ C^{W_2}_{12}\left(a_{\mbox{\footnotesize{$\MSbar$}}},
\alpha_{\mbox{\footnotesize{$\MSbar$}}}\right) 
\beta\left(a_{\mbox{\footnotesize{$\MSbar$}}}\right)
\frac{\partial ~}{\partial a_{\mbox{\footnotesize{$\MSbar$}}}}
\ln C^{W_2}_{11}\left(a_{\mbox{\footnotesize{$\MSbar$}}},
\alpha_{\mbox{\footnotesize{$\MSbar$}}}\right) \right. \nonumber \\
&& \left. ~+
C^{W_2}_{12}\!\left(a_{\mbox{\footnotesize{$\MSbar$}}},
\alpha_{\mbox{\footnotesize{$\MSbar$}}}\right) 
\alpha_{\mbox{\footnotesize{$\MSbar$}}}
\gamma^{\mbox{\footnotesize{$\MSbar$}}}_\alpha
\left(a_{\mbox{\footnotesize{$\MSbar$}}},
\alpha_{\mbox{\footnotesize{$\MSbar$}}}\right) \right. \nonumber \\
&& \left. ~~~~\times 
\frac{\partial ~}{\partial \alpha_{\mbox{\footnotesize{$\MSbar$}}}}
\ln C^{W_2}_{11}\left(a_{\mbox{\footnotesize{$\MSbar$}}},
\alpha_{\mbox{\footnotesize{$\MSbar$}}}\right) \right] 
\left[ C^{W_2}_{22}\left(a_{\mbox{\footnotesize{$\MSbar$}}},
\alpha_{\mbox{\footnotesize{$\MSbar$}}}\right)
\right]^{-1} 
\end{eqnarray}
and
\begin{eqnarray}
\gamma^{W_2}_{22,\mbox{\footnotesize{RI$^\prime$/SMOM}}}
\left(a_{\mbox{\footnotesize{RI$^\prime$}}},
\alpha_{\mbox{\footnotesize{RI$^\prime$}}}\right) &=&
\gamma^{W_2}_{22,\mbox{\footnotesize{$\MSbar$}}}
\left(a_{\mbox{\footnotesize{$\MSbar$}}}\right) ~-~
\beta\left(a_{\mbox{\footnotesize{$\MSbar$}}}\right)
\frac{\partial ~}{\partial a_{\mbox{\footnotesize{$\MSbar$}}}}
\ln C^{W_2}_{22}\left(a_{\mbox{\footnotesize{$\MSbar$}}},
\alpha_{\mbox{\footnotesize{$\MSbar$}}}\right) \nonumber \\
&& -~ \alpha_{\mbox{\footnotesize{$\MSbar$}}}
\gamma^{\mbox{\footnotesize{$\MSbar$}}}_\alpha
\left(a_{\mbox{\footnotesize{$\MSbar$}}},
\alpha_{\mbox{\footnotesize{$\MSbar$}}}\right)
\frac{\partial ~}{\partial \alpha_{\mbox{\footnotesize{$\MSbar$}}}}
\ln C^{W_2}_{22}\left(a_{\mbox{\footnotesize{$\MSbar$}}},
\alpha_{\mbox{\footnotesize{$\MSbar$}}}\right)
\end{eqnarray}
where to avoid confusion we have labelled the scheme the variables are in
explicitly. We have used the designation RI${}^\prime$ for the variables on the
left side of the equations as we use the definitions of \cite{24} which were 
derived from renormalizing the $2$-point functions of all the fields. With 
these definitions we have computed the conversion function matrix explicitly to
two loops and find\footnote{These together with the anomalous dimensions are 
also included in the attached electronic file.} 
\begin{eqnarray}
C^{W_2}_{11}(a,\alpha) &=& 1 ~+~ \left[ ( 36 \alpha - 42 ) \psi^\prime(\third) 
+ ( 28 - 24 \alpha ) \pi^2 - 27 \alpha + 459 \right] \frac{C_F a}{81} 
\nonumber \\
&& +~ \left[ \left[ 
( 5184 \alpha^2 - 12096 \alpha - 4608 ) (\psi^\prime(\third))^2 
+ ( 16128 \alpha - 6912 \alpha^2 + 6144 ) \psi^\prime(\third) \pi^2
\right. \right. \nonumber \\
&& \left. \left. ~~~~~
+~ ( 328536 \alpha - 13608 \alpha^2 - 613656 ) \psi^\prime(\third) 
- ( 5022 + 1944 \alpha ) \psi^{\prime\prime\prime}(\third) 
\right. \right. \nonumber \\
&& \left. \left. ~~~~~
+~ ( 1259712 \alpha - 5248800 ) s_2(\pisix) 
+ ( 10497600 - 2519424 \alpha ) s_2(\pitwo) 
\right. \right. \nonumber \\
&& \left. \left. ~~~~~
+~ ( 8748000 - 2099520 \alpha ) s_3(\pisix) 
+ ( 1679616 \alpha - 6998400 ) s_3(\pitwo) 
\right. \right. \nonumber \\
&& \left. \left. ~~~~~
+~ ( 2304 \alpha^2 - 192 \alpha + 11344 ) \pi^4 
+ ( 9072 \alpha^2 - 219024 \alpha + 409104 ) \pi^2 
\right. \right. \nonumber \\
&& \left. \left. ~~~~~
+~ 7290 \alpha^2 - 90882 \alpha + 107568
+ ( 34992 \alpha - 40824 ) \Sigma
\right. \right. \nonumber \\
&& \left. \left. ~~~~~
+~ ( 104976 \alpha + 559872 ) \zeta(3)
+ ( 8748 \alpha - 36450 ) \frac{\ln^2(3) \pi}{\sqrt{3}}
\right. \right. \nonumber \\
&& \left. \left. ~~~~~
+~ ( 437400 - 104976 \alpha ) \frac{\ln(3) \pi}{\sqrt{3}}
+ ( 39150 - 9396 \alpha ) \frac{\pi^3}{\sqrt{3}}
\right] C_F \right. \nonumber \\
&& ~~~~~+ \left. \left[ 
5832 (\psi^\prime(\third))^2 
- 7776 \psi^\prime(\third) \pi^2 
+ ( 11664 \alpha^2 - 39366 \alpha + 99468 ) \psi^\prime(\third) 
\right. \right. \nonumber \\
&& \left. \left. ~~~~~~~~~
+~ ( 5103 - 972 \alpha ) \psi^{\prime\prime\prime}(\third) 
+ ( 2519424 - 629856 \alpha ) s_2(\pisix) 
\right. \right. \nonumber \\
&& \left. \left. ~~~~~~~~~
+~ ( 1259712 \alpha - 5038848 ) s_2(\pitwo) 
+ ( 1049760 \alpha - 4199040 ) s_3(\pisix) 
\right. \right. \nonumber \\
&& \left. \left. ~~~~~~~~~
+~ ( 3359232 - 839808 \alpha ) s_3(\pitwo) 
+ ( 2592 \alpha - 11016 ) \pi^4 
\right. \right. \nonumber \\
&& \left. \left. ~~~~~~~~~
-~ ( 7776 \alpha^2 - 26244 \alpha + 66312 ) \pi^2 
- 8748 \alpha^2 - 34992 \alpha + 1759644
\right. \right. \nonumber \\
&& \left. \left. ~~~~~~~~~
+~ ( 17496 \alpha - 34992 ) \Sigma
+ ( 26244 \alpha - 734832 ) \zeta(3)
\right. \right. \nonumber \\
&& \left. \left. ~~~~~~~~~
+~ ( 17496 - 4374 \alpha ) \frac{\ln^2(3) \pi}{\sqrt{3}}
+ ( 52488 \alpha - 209952 ) \frac{\ln(3) \pi}{\sqrt{3}}
\right. \right. \nonumber \\
&& \left. \left. ~~~~~~~~~
+~ ( 4698 \alpha - 18792 ) \frac{\pi^3}{\sqrt{3}}
\right] C_A \right. \nonumber \\
&& ~~~~~+~ \left. \left[ 
46656 \psi^\prime(\third) 
- 31104 \pi^2 
- 642168
\right] T_F \Nf 
\right] \frac{C_F a^2}{26244} ~+~ O(a^3)
\end{eqnarray} 
\begin{eqnarray}
C^{W_2}_{12}(a,\alpha) &=& \left[ 24 \psi^\prime(\third) 
- 16 \pi^2 - 54 \alpha - 297 \right] \frac{C_F a}{162} 
\nonumber \\
&& +~ \left[ \left[ 
( 13824 \alpha + 35136 ) (\psi^\prime(\third))^2 
- ( 19440 \alpha^2 + 18432 \alpha + 46848 ) \psi^\prime(\third) \pi^2
\right. \right. \nonumber \\
&& \left. \left. ~~~~~
-~ ( 342144 \alpha + 6912 \alpha^2 - 824904 ) \psi^\prime(\third) 
+ 18792 \psi^{\prime\prime\prime}(\third) 
\right. \right. \nonumber \\
&& \left. \left. ~~~~~
+~ ( 10917504 - 1679616 \alpha ) s_2(\pisix) 
+ ( 3359232 - 21835008 \alpha ) s_2(\pitwo) 
\right. \right. \nonumber \\
&& \left. \left. ~~~~~
+~ ( 2799360 - 18195840 \alpha ) s_3(\pisix) 
+ ( 14556672 - 2239488 \alpha ) s_3(\pitwo) 
\right. \right. \nonumber \\
&& \left. \left. ~~~~~
+~ ( 6144 \alpha - 34496 ) \pi^4 
+ ( 12960 \alpha^2 + 228096 \alpha - 549936 ) \pi^2 
\right. \right. \nonumber \\
&& \left. \left. ~~~~~
+~ 37908 \alpha^2 + 76788 \alpha - 188892
+ 46656 \Sigma
\right. \right. \nonumber \\
&& \left. \left. ~~~~~
-~ ( 139968 \alpha + 1609632 ) \zeta(3)
+ ( 75816 - 11664 \alpha ) \frac{\ln^2(3) \pi}{\sqrt{3}}
\right. \right. \nonumber \\
&& \left. \left. ~~~~~
+~ ( 139968 \alpha - 909792 ) \frac{\ln(3) \pi}{\sqrt{3}}
+ ( 12528 \alpha - 81432 ) \frac{\pi^3}{\sqrt{3}}
\right] C_F \right. \nonumber \\
&& ~~~~~+ \left. \left[ 
31104 \psi^\prime(\third) \pi^2 
- 23328 (\psi^\prime(\third))^2 
+ ( 107892 \alpha - 5832 \alpha^2 - 122472 ) \psi^\prime(\third) 
\right. \right. \nonumber \\
&& \left. \left. ~~~~~~~~~
+~ ( 972 \alpha - 9720 ) \psi^{\prime\prime\prime}(\third) 
+ ( 944784 \alpha - 4513968 ) s_2(\pisix) 
\right. \right. \nonumber \\
&& \left. \left. ~~~~~~~~~
+~ ( 9027936 - 1889568 \alpha ) s_2(\pitwo) 
+ ( 7523280 - 1574640 \alpha ) s_3(\pisix) 
\right. \right. \nonumber \\
&& \left. \left. ~~~~~~~~~
+~ ( 1259712 \alpha - 6018624 ) s_3(\pitwo) 
+ ( 15552 - 2592 \alpha ) \pi^4 
\right. \right. \nonumber \\
&& \left. \left. ~~~~~~~~~
+~ ( 3888 \alpha^2 - 71928 \alpha + 81648 ) \pi^2 
- 21870 \alpha^2 - 113724 \alpha - 2488482
\right. \right. \nonumber \\
&& \left. \left. ~~~~~~~~~
+~ 34992 \Sigma
+ ( 1277208 - 52488 \alpha ) \zeta(3)
\right. \right. \nonumber \\
&& \left. \left. ~~~~~~~~~
+~ ( 6561 \alpha - 31347 ) \frac{\ln^2(3) \pi}{\sqrt{3}}
+ ( 376164 - 78732 \alpha ) \frac{\ln(3) \pi}{\sqrt{3}}
\right. \right. \nonumber \\
&& \left. \left. ~~~~~~~~~
+~ ( 33669 - 7047 \alpha ) \frac{\pi^3}{\sqrt{3}}
\right] C_A \right. \nonumber \\
&& ~~~~~+~ \left. \left[ 
29376 \pi^2 
- 44064 \psi^\prime(\third) 
+ 946080
\right] T_F \Nf 
\right] \frac{C_F a^2}{104976} ~+~ O(a^3)
\end{eqnarray} 
and
\begin{equation}
C^{W_2}_{22}(a,\alpha) ~=~ 1 ~+~ O(a^3) ~. 
\end{equation} 
For $SU(3)$ the numerical values are 
\begin{eqnarray}
C^{W_2}_{11}(a,\alpha) &=& 1 ~+~ [ 1.6390287 \alpha + 5.1248369 ] a \nonumber \\
&& +~ [ 6.5108151 \alpha^2 + 23.1781730 \alpha + 132.6228486 - 12.1458110 \Nf ]
a^2 ~+~ O(a^3) \nonumber \\
C^{W_2}_{12}(a,\alpha) &=& -~ [ 0.4444444 \alpha + 1.7499534 ] a \nonumber \\
&& -~ [ 2.1301455 \alpha^2 + 7.2772405 \alpha + 49.1967063 - 5.0243682 \Nf ]
a^2 ~+~ O(a^3) \nonumber \\
C^{W_2}_{22}(a,\alpha) &=& 1 ~+~ O(a^3) ~.
\end{eqnarray} 
As the operator $\partial W_2$ and the vector current are in effect equivalent 
then both their anomalous dimensions and conversion functions are the same. For
the latter case this is the $22$ element of both matrices. 

With these values we can now make a comparison with similar functions in the
RI${}^\prime$ scheme for the level $W_2$. However, in order to do this we need
to define the conversion functions from the renormalization constants. This is
not as straightforward for RI${}^\prime$ since the momentum configuration
defining the scheme does not access the off diagonal elements of the full 
operator mixing matrix. Given that we have chosen to define our operator basis 
in such a way that this matrix is triangular it is possible to define and 
compare the diagonal elements of the conversion function matrix. As noted 
for RI${}^\prime$ only the diagonal elements of the renormalization constant 
matrix exist and so we define the two conversion functions of interest as
\begin{equation}
\tilde{C}^{W_2}_{11}(a,\alpha) ~=~
\frac{Z^{W_2}_{11,\mbox{\footnotesize{RI$^\prime$}}}}
{Z^{W_2}_{11,\mbox{\footnotesize{$\MSbar$}}}} ~~~,~~~ 
\tilde{C}^{W_2}_{22}(a,\alpha) ~=~
\frac{Z^{W_2}_{22,\mbox{\footnotesize{RI$^\prime$}}}}
{Z^{W_2}_{22,\mbox{\footnotesize{$\MSbar$}}}} ~.  
\end{equation}
The explicit forms were given in \cite{24,25} and in the Landau gauge we have 
the $SU(3)$ numerical values 
\begin{eqnarray}
\tilde{C}^{W_2}_{11}(a,0) &=& 1 ~+~ 4.5925926 a ~+~ [ 119.8268158 
- 10.9794239 \Nf ] a^2 ~+~ O(a^3) \nonumber \\
\tilde{C}^{W_2}_{22}(a,0) &=& 1 ~+~ O(a^3) ~. 
\end{eqnarray} 
Clearly the numerical values of the one and two loop corrections for the
RI${}^\prime$ scheme are each about $10\%$ smaller than those for the
RI${}^\prime$/SMOM scheme for $\Nf$~$=$~$2$ and $3$. However, it is not 
entirely clear whether this is a proper comparison in the sense that one is not
necessarily comparing with respect to the same tensor basis. Equally the mixing
matrices are not truly the same as there can be no off-diagonal element for
RI${}^\prime$. Indeed it is probable that the convergence of the 
RI${}^\prime$/SMOM scheme result could be improved by a different choice of
basis tensors or exploit the freedom in how one defines the RI${}^\prime$/SMOM 
scheme for this operator. That aside one does at least have the explicit forms 
of the amplitudes at the symmetric subtraction point at two loops in order to 
assist with lattice matching.

One benefit of the explicit forms of the conversion functions is that we can
deduce the {\em three} loop RI${}^\prime$/SMOM scheme anomalous dimensions in 
the Landau gauge using the three loop $\MSbar$ expressions of
(\ref{anomdimms}). We have 
\begin{eqnarray}
\left. \frac{}{} \gamma^{W_2}_{11}(a,0) 
\right|_{\mbox{\footnotesize{RI$^\prime$/SMOM}}} &=& 
\frac{8}{3} C_F a 
\nonumber \\
&& +~ \left[ \left[ 308 \pi^2 - 462 \psi^\prime(\third) + 8433 \right] C_A ~-~
1008 C_F \right. \nonumber \\
&& \left. ~~~~+~ \left[ 168 \psi^\prime(\third) - 112 \pi^2 - 2988 \right]
T_F \Nf \right] \frac{C_F a^2}{243} ~+~ O(a^3) \nonumber \\
&& +~ \left[ 
\left[
64152 (\psi^\prime(\third))^2 
- 85536 \psi^\prime(\third) \pi^2
+ 862812 \psi^\prime(\third)
\right. \right. \nonumber \\
&& \left. \left. ~~~~~
+~ 56133 \psi^{\prime\prime\prime}(\third) 
+ 27713664 s_2(\pisix) 
- 55427328 s_2(\pitwo) 
\right. \right. \nonumber \\
&& \left. \left. ~~~~~
-~ 46189440 s_3(\pisix) 
+ 36951552 s_3(\pitwo) 
- 121176 \pi^4 
- 575208 \pi^2
\right. \right. \nonumber \\
&& \left. \left. ~~~~~
-~ 384912 \Sigma
- 7243344 \zeta(3)
+ 25273296
+ 192456 \frac{\ln^2(3) \pi}{\sqrt{3}}
\right. \right. \nonumber \\
&& \left. \left. ~~~~~
-~ 2309472 \frac{\ln(3) \pi}{\sqrt{3}}
- 206712 \frac{\pi^3}{\sqrt{3}}
\right] C_A^2 \right. \nonumber \\
&& \left. ~~~~+ \left[
119328 \psi^\prime(\third) \pi^2
- 89496 (\psi^\prime(\third))^2 
- 5901984 \psi^\prime(\third) 
\right. \right. \nonumber \\
&& \left. \left. ~~~~~~~~
-~ 55242 \psi^{\prime\prime\prime}(\third) 
- 57736800 s_2(\pisix) 
+ 115473600 s_2(\pitwo) 
\right. \right. \nonumber \\
&& \left. \left. ~~~~~~~~
+~ 96228000 s_3(\pisix) 
- 76982400 s_3(\pitwo) 
+ 107536 \pi^4 
\right. \right. \nonumber \\
&& \left. \left. ~~~~~~~~
+~ 3934656 \pi^2
- 449064 \Sigma
+ 3639168 \zeta(3)
- 4833270
\right. \right. \nonumber \\
&& \left. \left. ~~~~~~~~
-~ 400950 \frac{\ln^2(3) \pi}{\sqrt{3}}
+ 4811400 \frac{\ln(3) \pi}{\sqrt{3}}
+ 430650 \frac{\pi^3}{\sqrt{3}}
\right] C_A C_F \right. \nonumber \\
&& \left. ~~~~+ \left[
31104 (\psi^\prime(\third))^2 
- 23328 \psi^\prime(\third) \pi^2
+ 251424 \psi^\prime(\third) 
\right. \right. \nonumber \\
&& \left. \left. ~~~~~~~~
-~ 20412 \psi^{\prime\prime\prime}(\third) 
- 10077696 s_2(\pisix) 
+ 20155392 s_2(\pitwo) 
\right. \right. \nonumber \\
&& \left. \left. ~~~~~~~~
+~ 16796160 s_3(\pisix) 
- 13436928 s_3(\pitwo) 
+ 44064 \pi^4 
\right. \right. \nonumber \\
&& \left. \left. ~~~~~~~~
-~ 167616 \pi^2
+ 139968 \Sigma
+ 1259712 \zeta(3)
- 16603056
\right. \right. \nonumber \\
&& \left. \left. ~~~~~~~~
-~ 69984 \frac{\ln^2(3) \pi}{\sqrt{3}}
+ 839808 \frac{\ln(3) \pi}{\sqrt{3}}
+ 75168 \frac{\pi^3}{\sqrt{3}}
\right] C_A T_F \Nf \right. \nonumber \\
&& \left. ~~~~+ \left[
32544 (\psi^\prime(\third))^2 
- 43392 \psi^\prime(\third) \pi^2
+ 2227824 \psi^\prime(\third) 
\right. \right. \nonumber \\
&& \left. \left. ~~~~~~~~
+~ 20088 \psi^{\prime\prime\prime}(\third) 
+ 20995200 s_2(\pisix) 
- 41990400 s_2(\pitwo) 
\right. \right. \nonumber \\
&& \left. \left. ~~~~~~~~
-~ 34992000 s_3(\pisix) 
+ 27993600 s_3(\pitwo) 
- 39104 \pi^4 
\right. \right. \nonumber \\
&& \left. \left. ~~~~~~~~
-~ 1485216 \pi^2
+ 163296 \Sigma
- 559872 \zeta(3)
- 742608
\right. \right. \nonumber \\
&& \left. \left. ~~~~~~~~
+\, 145800 \frac{\ln^2(3) \pi}{\sqrt{3}}
- 1749600 \frac{\ln(3) \pi}{\sqrt{3}}
- 156600 \frac{\pi^3}{\sqrt{3}}
\right] C_F T_F \Nf \right. \nonumber \\
&& \left. ~~~~+ \left[
124416 \pi^2 
- 186624 \psi^\prime(\third) 
+ 2423520 
\right] T_F^2 \Nf^2 \right. \nonumber \\
&& \left. ~~~~+ \left[
1679616 \zeta(3) 
- 90720 
\right] C_F^2 \right] \frac{C_F a^3}{39366} ~+~ O(a^4) 
\end{eqnarray} 
\begin{eqnarray}
\left. \frac{}{} \gamma^{W_2}_{12}(a,0) 
\right|_{\mbox{\footnotesize{RI$^\prime$/SMOM}}} &=& 
-~ \frac{4}{3} C_F a 
\nonumber \\
&& +~ \left[ \left[ 264 \psi^\prime(\third) - 176 \pi^2 - 6651 \right] C_A ~+~
\left[ 144 \psi^\prime(\third) - 96 \pi^2 - 288 \right] C_F \right. 
\nonumber \\
&& \left. ~~~~+~ \left[ 64 \pi^2 - 96 \psi^\prime(\third) + 2340 \right]
T_F \Nf \right] \frac{C_F a^2}{486} ~+~ O(a^3) \nonumber \\
&& +~ \left[ 
\left[
342144 \psi^\prime(\third) \pi^2
- 256608 (\psi^\prime(\third))^2 
- 1082808 \psi^\prime(\third) 
\right. \right. \nonumber \\
&& \left. \left. ~~~~~
-~ 106920 \psi^{\prime\prime\prime}(\third) 
- 49653648 s_2(\pisix) 
+ 99307296 s_2(\pitwo) 
\right. \right. \nonumber \\
&& \left. \left. ~~~~~
+~ 82756080 s_3(\pisix) 
- 66204864 s_3(\pitwo) 
+ 171072 \pi^4 
+ 721872 \pi^2
\right. \right. \nonumber \\
&& \left. \left. ~~~~~
+~ 384912 \Sigma
+ 12369672 \zeta(3)
- 37423134
- 344817 \frac{\ln^2(3) \pi}{\sqrt{3}}
\right. \right. \nonumber \\
&& \left. \left. ~~~~~
+~ 4137804 \frac{\ln(3) \pi}{\sqrt{3}}
+ 370359 \frac{\pi^3}{\sqrt{3}}
\right] C_A^2 \right. \nonumber \\
&& \left. ~~~~+ \left[
477504 (\psi^\prime(\third))^2 
- 636672 \psi^\prime(\third) \pi^2
+ 7978176 \psi^\prime(\third) 
\right. \right. \nonumber \\
&& \left. \left. ~~~~~~~~
+~ 204768 \psi^{\prime\prime\prime}(\third) 
+ 117993024 s_2(\pisix) 
- 235986048 s_2(\pitwo) 
\right. \right. \nonumber \\
&& \left. \left. ~~~~~~~~
-~ 196655040 s_3(\pisix) 
+ 157324032 s_3(\pitwo) 
- 333824 \pi^4 
\right. \right. \nonumber \\
&& \left. \left. ~~~~~~~~
-~ 5318784 \pi^2
+ 653184 \Sigma
- 11897280 \zeta(3)
+ 367416
\right. \right. \nonumber \\
&& \left. \left. ~~~~~~~~
+~ 819396 \frac{\ln^2(3) \pi}{\sqrt{3}}
- 9832752 \frac{\ln(3) \pi}{\sqrt{3}}
- 880092 \frac{\pi^3}{\sqrt{3}}
\right] C_A C_F \right. \nonumber \\
&& \left. ~~~~+ \left[
93312 (\psi^\prime(\third))^2 
- 124416 \psi^\prime(\third) \pi^2
- 150336 \psi^\prime(\third) 
\right. \right. \nonumber \\
&& \left. \left. ~~~~~~~~
+~ 38880 \psi^{\prime\prime\prime}(\third) 
+ 18055872 s_2(\pisix) 
- 36111744 s_2(\pitwo) 
\right. \right. \nonumber \\
&& \left. \left. ~~~~~~~~
-~ 30093120 s_3(\pisix) 
+ 24074496 s_3(\pitwo) 
- 62208 \pi^4 
\right. \right. \nonumber \\
&& \left. \left. ~~~~~~~~
+~ 100224 \pi^2
- 139968 \Sigma
- 1749600 \zeta(3)
+ 24312312
\right. \right. \nonumber \\
&& \left. \left. ~~~~~~~~
+\, 125388 \frac{\ln^2(3) \pi}{\sqrt{3}}
- 1504656 \frac{\ln(3) \pi}{\sqrt{3}}
- 134676 \frac{\pi^3}{\sqrt{3}}
\right] C_A T_F \Nf \right. \nonumber \\
&& \left. ~~~~+ \left[
208896 \psi^\prime(\third) \pi^2
- 156672 (\psi^\prime(\third))^2 
- 3297024 \psi^\prime(\third) 
\right. \right. \nonumber \\
&& \left. \left. ~~~~~~~~
-~ 75168 \psi^{\prime\prime\prime}(\third) 
- 43670016 s_2(\pisix) 
+ 87340032 s_2(\pitwo) 
\right. \right. \nonumber \\
&& \left. \left. ~~~~~~~~
+~ 72783360 s_3(\pisix) 
- 58226688 s_3(\pitwo) 
+ 130816 \pi^4 
\right. \right. \nonumber \\
&& \left. \left. ~~~~~~~~
+~ 2198016 \pi^2
- 186624 \Sigma
+ 3079296 \zeta(3)
+ 4039632
\right. \right. \nonumber \\
&& \left. \left. ~~~~~~~~
-\, 303264 \frac{\ln^2(3) \pi}{\sqrt{3}}
+ 3639168 \frac{\ln(3) \pi}{\sqrt{3}}
+ 325728 \frac{\pi^3}{\sqrt{3}}
\right] C_F T_F \Nf \right. \nonumber \\
&& \left. ~~~~+ \left[
176256 \psi^\prime(\third)
- 117504 \pi^2 
- 3494016 
\right] T_F^2 \Nf^2 \right. \nonumber \\
&& \left. ~~~~+ \left[
138240 \psi^\prime(\third) \pi^2
- 103680 (\psi^\prime(\third))^2 
+ 1537056 \psi^\prime(\third) 
\right. \right. \nonumber \\
&& \left. \left. ~~~~~~~~
-~ 34992 \psi^{\prime\prime\prime}(\third) 
- 1679616 s_2(\pisix) 
+ 3359232 s_2(\pitwo) 
\right. \right. \nonumber \\
&& \left. \left. ~~~~~~~~
+~ 2799360 s_3(\pisix) 
- 2239488 s_3(\pitwo) 
+ 47232 \pi^4 
\right. \right. \nonumber \\
&& \left. \left. ~~~~~~~~
-~ 1024704 \pi^2
+ 139968 \Sigma
- 1399680 \zeta(3)
+ 729648
\right. \right. \nonumber \\
&& \left. \left. ~~~~~~~~
-\, 11664 \frac{\ln^2(3) \pi}{\sqrt{3}}
+ 139968 \frac{\ln(3) \pi}{\sqrt{3}}
+ 12528 \frac{\pi^3}{\sqrt{3}}
\right] C_F^2 \right] \frac{C_F a^3}{157464} \nonumber \\
&& +~ O(a^4) 
\end{eqnarray} 
and 
\begin{eqnarray}
\left. \frac{}{} \gamma^{W_2}_{22}(a,0) 
\right|_{\mbox{\footnotesize{RI$^\prime$/SMOM}}} &=&  O(a^4) ~. 
\end{eqnarray} 
Again the final expression is the same as that for the vector operator. In
numerical form we have 
\begin{eqnarray}
\left. \gamma^{W_2}_{11}(a,0) \right|_{\mbox{\footnotesize{RI$^\prime$/SMOM}}} 
&=& 3.5555556 a ~+~ \left[ 104.7024244 - 6.5770518 \Nf \right] a^2 \nonumber \\
&& +~ \left[ 4010.9803829 - 624.8817671 \Nf + 14.9653337 \Nf^2 \right] 
a^3 ~+~ O(a^4) \nonumber \\
\left. \gamma^{W_2}_{12}(a,0) \right|_{\mbox{\footnotesize{RI$^\prime$/SMOM}}} 
&=& -~ 1.7777778 a ~-~ \left[ 46.3028612 - 2.7468825 \Nf \right] a^2 
\nonumber \\
&& -~ \left[ 1692.0143513 - 265.3339715 \Nf + 6.0846171 \Nf^2 \right] a^3 ~+~ 
O(a^4) \nonumber \\
\left. \gamma^{W_2}_{22}(a,0) \right|_{\mbox{\footnotesize{RI$^\prime$/SMOM}}} 
&=& O(a^4)
\end{eqnarray} 
for $SU(3)$. 

\sect{Discussion.}

We have provided the complete set of amplitudes at two loops for the second
moment of the Wilson operators inserted in a quark $2$-point function in both 
the $\MSbar$ and RI${}^\prime$/SMOM renormalization schemes. This is not as 
straightforward a task in comparison with earlier work to this level, 
\cite{28,29}, as there is mixing with a total derivative operator. One of the 
aims of the original RI${}^\prime$/SMOM scheme was that the convergence of the 
conversion functions between these two schemes would improve compared with the 
RI${}^\prime$ scheme. Indeed for the quark currents that appears to be the 
case. However, for $W_2$ if anything the RI${}^\prime$ result seems to be 
converging marginally quicker. Though it is not completely clear if this is
really an appropriate comparison. This is because the operator mixing simply 
does not arise in the RI${}^\prime$ case due to the very nature of the momentum 
configuration used for the Green's function. Indeed given the lack of
multiplicative renormalizability and hence mixing, it is not entirely clear 
what the status of the RI${}^\prime$ scheme is for $W_2$. It may be that 
RI${}^\prime$/SMOM is the only proper scheme to use of the two. Moreover, 
RI${}^\prime$/SMOM should not suffer the infrared issue associated with 
RI${}^\prime$, \cite{27}. However, if one was concerned about improving the 
convergence of the conversion function it might be possible to exploit the 
freedom one has in actually defining the scheme. For the operator $W_2$ we used
the channel $1$ and $2$ amplitudes as the basis for the renormalization 
conditions where these channels are defined with respect to a choice of basis 
tensors. This basis is by no means unique as one could equally choose another 
basis and hence use the analogous channels $1$ and $2$ there. As a variation 
one could instead use a different combination of amplitudes which equates to 
projecting by a linear combination of basis tensors. The advantage of this is 
that one would incorporate more information about the operator within the 
Green's function which is encoded in the other amplitudes. Whilst 
mathematically this is not inequivalent to a different choice of basis tensors,
it would avoid the tedious reworking of the construction of the projection 
matrix and thereafter the running of the underlying computer algebra 
programmes. This is one of the reasons why we have provided the $\MSbar$ 
results so that an interested reader has the information for whatever scheme 
definition variation one might conceive. As was noted in \cite{31} the 
renormalization of the next moment of the set of Wilson operators involves that
of $W_2$, because of mixing into a total derivative of $W_2$ itself. Therefore,
our results for the anomalous dimensions, amplitudes and conversion functions 
for level $W_2$ will provide important checks on the $n$~$=$~$3$ Wilson
operator renormalization in RI${}^\prime$/SMOM, \cite{53}.  

\vspace{1cm}
\noindent
{\bf Acknowledgement.} The author thanks Dr. P.E.L. Rakow for useful 
discussions.

\clearpage

{\begin{table}[ht]
\begin{center}
\begin{tabular}{|c||r|r|r|r|r|}
\hline
$a^{(1)}_n$ & $\left. c^{W_2,(1)}_{(1)\,n} \right|_{\MSbars}$ 
& $\left. c^{W_2,(1)}_{(2)\,n} \right|_{\MSbars}$ 
& $\left. c^{W_2,(1)}_{(3)\,n} \right|_{\MSbars}$ 
& $\left. c^{W_2,(1)}_{(4)\,n} \right|_{\MSbars}$ 
& $\left. c^{W_2,(1)}_{(5)\,n} \right|_{\MSbars}$ \\
\hline
$1$ & $- 11/6$ & $23/6$ & $16/27$ & $71/27$ & $100/27$ \\
$\pi^2 \alpha$ & $0$ & $- 8/27$ & $- 16/81$ & $- 32/81$ & $- 64/81$ \\
$\alpha$ & $- 1/3$ & $- 5/3$ & $- 8/9$ & $- 16/9$ & $- 26/9$ \\
$\pi^2$ & $- 8/81$ & $20/81$ & $- 16/243$ & $64/243$ & $80/243$ \\
$\psi^\prime(1/3)$ & $4/27$ & $- 10/27$ & $8/81$ & $- 32/81$ & $- 40/81$  \\
$\psi^\prime(1/3) \alpha$ & $0$ & $4/9$ & $8/27$ & $16/27$ & $32/27$ \\
\hline
\end{tabular}
\end{center}
\begin{center}
{Table $1$. Coefficients of $C_F$ for one loop $\MSbar$ $W_2$ amplitudes.}
\end{center}
\end{table}}

\vspace{3cm}
{\begin{table}[ht]
\begin{center}
\begin{tabular}{|c||r|r|r|r|r|r|r|r|r|r|}
\hline
$a^{(1)}_n$ & $\left. c^{W_2,(1)}_{(6)\,n} \right|_{\MSbars}$ 
& $\left. c^{W_2,(1)}_{(7)\,n} \right|_{\MSbars}$
& $\left. c^{W_2,(1)}_{(8)\,n} \right|_{\MSbars}$ 
& $\left. c^{W_2,(1)}_{(9)\,n} \right|_{\MSbars}$ 
& $\left. c^{W_2,(1)}_{(10)\,n} \right|_{\MSbars}$ \\
\hline
$1$ & $- 28/27$ & $37/27$ & $128/27$ & $- 1/3$ & $1/3$ \\
$\pi^2 \alpha$ & $16/81$ & $- 16/81$ & $- 32/81$ & $0$ & $0$ \\
$\alpha$ & $14/9$ & $- 2/9$ & $- 16/9$ & $0$ & $0$ \\
$\pi^2$ & $- 80/243$ & $8/243$ & $160/243$ & $0$ & $8/27$ \\
$\psi^\prime(1/3)$ & $40/81$ & $- 4/81$ & $- 80/81$ & $0$ & $- 4/9$ \\
$\psi^\prime(1/3) \alpha$ & $- 8/27$ & $8/27$ & $16/27$ & $0$ & $0$ \\
\hline
\end{tabular}
\end{center}
\begin{center}
{Table $2$. Coefficients of $C_F$ for one loop $\MSbar$ $W_2$ amplitudes
continued.}
\end{center}
\end{table}}

\vspace{3cm}
{\begin{table}[ht]
\begin{center}
\begin{tabular}{|c||r|r|r|r|r|}
\hline
$a^{(1)}_n$ & $\left. c^{\partial W_2,(1)}_{(1)\,n} \right|_{\MSbars}$ 
& $\left. c^{\partial W_2,(1)}_{(3)\,n} \right|_{\MSbars}$ 
& $\left. c^{\partial W_2,(1)}_{(4)\,n} \right|_{\MSbars}$ 
& $\left. c^{\partial W_2,(1)}_{(5)\,n} \right|_{\MSbars}$ 
& $\left. c^{\partial W_2,(1)}_{(9)\,n} \right|_{\MSbars}$ \\
\hline
$1$ & $2$ & $16/3$ & $4$ & $8/3$ & $0$ \\
$\pi^2 \alpha$ & $- 8/27$ & $- 16/27$ & $- 16/27$ & $- 16/27$ & $0$ \\
$\alpha$ & $- 2$ & $- 8/3$ & $- 2$ & $- 4/3$ & $0$ \\
$\pi^2$ & $4/27$ & $16/27$ & $- 8/27$ & $0$ & $8/27$ \\
$\psi^\prime(1/3)$ & $- 2/9$ & $- 8/9$ & $- 4/9$ & $0$ & $- 4/9$  \\
$\psi^\prime(1/3) \alpha$ & $4/9$ & $8/9$ & $8/9$ & $8/9$ & $0$ \\
\hline
\end{tabular}
\end{center}
\begin{center}
{Table $3$. Coefficients of $C_F$ for one loop $\MSbar$ $\partial W_2$ 
amplitudes.}
\end{center}
\end{table}}

\clearpage

{\begin{table}[htb]
\begin{center}
\begin{tabular}{|c||r|r|r|r|r|}
\hline
$a^{(21)}_n$ & $\left. c^{W_2,(21)}_{(1)\,n} \right|_{\MSbars}$ 
& $\left. c^{W_2,(21)}_{(2)\,n} \right|_{\MSbars}$ 
& $\left. c^{W_2,(21)}_{(3)\,n} \right|_{\MSbars}$ 
& $\left. c^{W_2,(21)}_{(4)\,n} \right|_{\MSbars}$ 
& $\left. c^{W_2,(21)}_{(5)\,n} \right|_{\MSbars}$ \\
\hline
$1$ & $730/81$ & $- 1937/162$ & $8/27$ & $- 8$ & $- 308/27$ \\
$\pi^2 \alpha$ & $0$ & $0$ & $0$ & $0$ & $0$ \\
$\alpha$ & $0$ & $0$ & $0$ & $0$ & $0$ \\
$\pi^2$ & $68/243$ & $- 220/243$ & $80/243$ & $- 248/243$ & $32/27$ \\
$\psi^\prime(1/3)$ & $- 34/81$ & $110/81$ & $- 40/81$ & $124/81$ & $32/27$ \\
$\psi^\prime(1/3) \alpha$ & $0$ & $0$ & $0$ & $0$ & $0$ \\
\hline
\end{tabular}
\end{center}
\begin{center}
{Table $4$. Coefficients of $C_F T_F \Nf$ for two loop $\MSbar$ $W_2$ 
amplitudes.}
\end{center}
\end{table}}

\vspace{3cm}
{\begin{table}[htb]
\begin{center}
\begin{tabular}{|c||r|r|r|r|r|}
\hline
$a^{(21)}_n$ & $\left. c^{W_2,(21)}_{(6)\,n} \right|_{\MSbars}$ 
& $\left. c^{W_2,(21)}_{(7)\,n} \right|_{\MSbars}$ 
& $\left. c^{W_2,(21)}_{(8)\,n} \right|_{\MSbars}$ 
& $\left. c^{W_2,(21)}_{(9)\,n} \right|_{\MSbars}$ 
& $\left. c^{W_2,(21)}_{(10)\,n} \right|_{\MSbars}$ \\
\hline
$1$ & $76/27$ & $- 44/9$ & $- 472/27$ & $32/27$ & $- 32/27$ \\
$\pi^2 \alpha$ & $0$ & $0$ & $0$ & $0$ & $0$ \\
$\alpha$ & $0$ & $0$ & $0$ & $0$ & $0$ \\
$\pi^2$ & $68/41$ & $- 56/243$ & $- 688/243$ & $8/81$ & $- 232/243$ \\
$\psi^\prime(1/3)$ & $- 32/27$ & $26/81$ & $344/81$ & $- 4/27$ & $116/81$ \\
$\psi^\prime(1/3) \alpha$ & $0$ & $0$ & $0$ & $0$ & $0$ \\
\hline
\end{tabular}
\end{center}
\begin{center}
{Table $5$. Coefficients of $C_F T_F \Nf$ for two loop $\MSbar$ $W_2$ 
amplitudes continued.}
\end{center}
\end{table}}

\vspace{3cm}
{\begin{table}[htb]
\begin{center}
\begin{tabular}{|c||r|r|r|r|r|}
\hline
$a^{(21)}_n$ & $\left. c^{\partial W_2,(21)}_{(1)\,n} \right|_{\MSbars}$ 
& $\left. c^{\partial W_2,(21)}_{(3)\,n} \right|_{\MSbars}$ 
& $\left. c^{\partial W_2,(21)}_{(4)\,n} \right|_{\MSbars}$ 
& $\left. c^{\partial W_2,(21)}_{(5)\,n} \right|_{\MSbars}$ 
& $\left. c^{\partial W_2,(21)}_{(9)\,n} \right|_{\MSbars}$ \\
\hline
$1$ & $- 53/18$ & $- 464/27$ & $- 116/9$ & $- 232/27$ & $- 208/243$ \\
$\pi^2 \alpha$ & $0$ & $0$ & $0$ & $0$ & $0$ \\
$\alpha$ & $0$ & $0$ & $0$ & $0$ & $0$ \\
$\pi^2$ & $- 152/243$ & $- 608/243$ & $- 304/243$ & $0$ & $0$ \\
$\psi^\prime(1/3)$ & $77/81$ & $304/81$ & $152/81$ & $0$ & $104/81$ \\
$\psi^\prime(1/3) \alpha$ & $0$ & $0$ & $0$ & $0$ & $0$ \\
\hline
\end{tabular}
\end{center}
\begin{center}
{Table $6$. Coefficients of $C_F T_F \Nf$ for two loop $\MSbar$ $\partial W_2$ 
amplitudes.}
\end{center}
\end{table}}

\clearpage

{\begin{table}[ht]
\begin{center}
\begin{tabular}{|c||r|r|r|r|r|}
\hline
$a^{(22)}_n$ & $\left. c^{W_2,(22)}_{(1)\,n} \right|_{\MSbars}$ 
& $\left. c^{W_2,(22)}_{(2)\,n} \right|_{\MSbars}$ 
& $\left. c^{W_2,(22)}_{(3)\,n} \right|_{\MSbars}$ 
& $\left. c^{W_2,(22)}_{(4)\,n} \right|_{\MSbars}$ 
& $\left. c^{W_2,(22)}_{(5)\,n} \right|_{\MSbars}$ \\
\hline
$1$ & $- 15361/648$ & $ 21445/648 $ & $- 188/81$ & $8351/324$ & $5879/162$ \\
$\pi^2 \alpha$ & $- 37/54$ & $17/54$ & $- 214/81$ & $67/81$ & $266/81$ \\
$\pi^4 \alpha$ & $- 2/81$ & $2/27$ & $- 8/243$ & $32/243$ & $76/243$ \\
$\zeta(3) \alpha$ & $- 1/2$ & $7/2$ & $- 2$ & $7/3$ & $6$ \\
$\Sigma \alpha$ & $0$ & $2/3$ & $4/9$ & $8/9$ & $16/9$ \\
$\alpha$ & $- 13/12$ & $- 107/12$ & $- 23/9$ & $- 125/18$ & $- 107/9$ \\
$\pi^2 \alpha^2$ & $1/27$ & $- 7/27$ & $- 10/81$ & $- 26/81$ & $- 58/81$ \\
$\alpha^2$ & $- 5/24$ & $- 5/3$ & $- 5/9$ & $- 55/36$ & $- 49/18$ \\
$\pi^2 \alpha^3$ & $0$ & $0$ & $0$ & $0$ & $0$ \\
$\alpha^3$ & $0$ & $0$ & $0$ & $0$ & $0$ \\
$\pi^2$ & $7/9$ & $- 425/243$ & $7724/729$ & $- 2384/729$ & $- 7774/729$ \\
$\pi^4$ & $4/27$ & $- 22/81$ & $200/243$ & $- 100/243$ & $- 68/81$ \\
$\zeta(3)$ & $73/6$ & $- 77/6$ & $98/3$ & $- 41/3$ & $- 34$ \\
$\Sigma$ & $1/3$ & $- 1$ & $4/9$ & $- 10/9$ & $- 8/9$ \\
$s_2(\pi/6)$ & $- 43$ & $53$ & $- 208$ & $70$ & $168$ \\
$s_2(\pi/6) \alpha$ & $9$ & $- 15$ & $24$ & $- 26$ & $- 72$ \\
$s_2(\pi/2)$ & $86$ & $- 106$ & $416$ & $- 140$ & $- 336$ \\
$s_2(\pi/2) \alpha$ & $- 18$ & $30$ & $- 48$ & $52$ & $144$ \\
$s_3(\pi/6)$ & $215/3$ & $- 265/3$ & $1040/3$ & $- 350/3$ & $- 280$ \\
$s_3(\pi/6) \alpha$ & $- 15$ & $25$ & $- 40$ & $130/3$ & $120$ \\
$s_3(\pi/2)$ & $- 172/3$ & $212/3$ & $- 832/3$ & $280/3$ & $224$ \\
$s_3(\pi/2) \alpha$ & $12$ & $- 20$ & $32$ & $- 104/3$ & $- 96$ \\
$\psi^\prime(1/3)$ & $- 7/6$ & $425/162$ & $- 3862/243$ & $1192/243$ & $3887/243$ \\
$\psi^\prime(1/3) \alpha$ & $37/36$ & $- 17/36$ & $107/27$ & $67/54$ & $- 133/27$ \\
$\psi^\prime(1/3) \alpha^2$ & $- 1/18$ & $7/18$ & $5/27$ & $13/27$ & $29/27$ \\
$\psi^\prime(1/3) \alpha^3$ & $0$ & $0$ & $0$ & $0$ & $0$ \\
$\psi^\prime(1/3) \pi^2 $ & $8/27$ & $0$ & $64/81$ & $16/81$ & $0$ \\
$(\psi^\prime(1/3))^2$ & $- 2/9$ & $0$ & $- 16/27$ & $- 4/27$ & $0$ \\
$\psi^{\prime\prime\prime}(1/3)$ & $- 5/54$ & $11/108$ & $- 11/27$ & $7/54$ & $17/54$ \\
$\psi^{\prime\prime\prime}(1/3) \alpha$ & $1/108$ & $- 1/36$ & $1/81$ & $- 4/81$ & $- 19/162$ \\
$\pi^3 \alpha/\sqrt{3}$ & $- 29/432$ & $145/1296$ & $- 29/162$ & $377/1944$ & $29/54$ \\
$\pi^3/\sqrt{3}$ & $1247/3888$ & $- 1537/3888$ & $377/243$ & $- 1015/1944$ & $- 203/162$ \\
$\pi \ln(3) \alpha/\sqrt{3}$ & $- 3/4$ & $5/4$ & $- 2$ & $13/6$ & $6$ \\
$\pi \ln(3)/\sqrt{3}$ & $43/12$ & $- 53/12$ & $52/3$ & $- 35/6$ & $- 14$ \\
$\pi (\ln(3))^2 \alpha/\sqrt{3}$ & $1/16$ & $- 5/48$ & $1/6$ & $- 13/72$ & $- 1/2$ \\
$\pi (\ln(3))^2/\sqrt{3}$ & $- 43/144$ & $53/144$ & $- 13/9$ & $35/72$ & $7/6$ \\
\hline
\end{tabular}
\end{center}
\begin{center}
{Table $7$. Coefficients of $C_F C_A$ for two loop $\MSbar$ $W_2$ amplitudes.}
\end{center}
\end{table}}

\clearpage

{\begin{table}[ht]
\begin{center}
\begin{tabular}{|c||r|r|r|r|r|}
\hline
$a^{(22)}_n$ & $\left. c^{W_2,(22)}_{(6)\,n} \right|_{\MSbars}$ 
& $\left. c^{W_2,(22)}_{(7)\,n} \right|_{\MSbars}$ 
& $\left. c^{W_2,(22)}_{(8)\,n} \right|_{\MSbars}$ 
& $\left. c^{W_2,(22)}_{(9)\,n} \right|_{\MSbars}$ 
& $\left. c^{W_2,(22)}_{(10)\,n} \right|_{\MSbars}$ \\
\hline
$1$ & $- 1637/162$ & $4375/324$ & $4430/81$ & $- 379/108$ & $ 379/108$ \\
$\pi^2 \alpha$ & $- 218/81$ & $- 127/81$ & $46/81$ & $7/27$ & $- 1/27$ \\
$\pi^4 \alpha$ & $- 52/243$ & $- 8/243$ & $32/243$ & $0$ & $- 4/81$ \\
$\zeta(3) \alpha$ & $- 6$ & $- 7/3$ & $2$ & $1/3$ & $- 1$ \\
$\Sigma \alpha$ & $- 4/9$ & $4/9$ & $8/9$ & $0$ & $0$ \\
$\alpha$ & $65/9$ & $- 1/18$ & $- 61/9$ & $- 1/6$ & $1/6$ \\
$\pi^2 \alpha^2$ & $22/81$ & $- 10/81$ & $- 26/81$ & $0$ & $2/27$ \\
$\alpha^2$ & $31/18$ & $1/36$ & $- 13/9$ & $0$ & $1/12$ \\
$\pi^2 \alpha^3$ & $0$ & $0$ & $0$ & $0$ & $0$ \\
$\alpha^3$ & $0$ & $0$ & $0$ & $0$ & $0$ \\
$\pi^2$ & $3346/729$ & $968/729$ & $- 6128/729$ & $- 284/81$ & $- 820/243$ \\
$\pi^4$ & $148/243$ & $40/243$ & $- 88/81$ & $- 44/243$ & $- 52/243$ \\
$\zeta(3)$ & $70/3$ & $19/3$ & $- 110/3$ & $- 5$ & $- 13/3$ \\
$\Sigma$ & $8/9$ & $- 2/9$ & $- 28/9$ & $- 2/9$ & $- 10/9$ \\
$s_2(\pi/6)$ & $- 128$ & $- 50$ & $208$ & $30$ & $38$ \\
$s_2(\pi/6) \alpha$ & $48$ & $14$ & $- 24$ & $- 2$ & $6$ \\
$s_2(\pi/2)$ & $256$ & $100$ & $- 416$ & $- 60$ & $- 76$ \\
$s_2(\pi/2) \alpha$ & $- 96$ & $- 28$ & $48$ & $4$ & $- 12$ \\
$s_3(\pi/6)$ & $640/3$ & $250/3$ & $- 1040/3$ & $50$ & $- 190/3$ \\
$s_3(\pi/6) \alpha$ & $- 80$ & $- 70/3$ & $40$ & $- 10/3$ & $- 10$ \\
$s_3(\pi/2)$ & $- 512/3$ & $- 200/3$ & $832/3$ & $40$ & $152/3$ \\
$s_3(\pi/2) \alpha$ & $64$ & $56/3$ & $- 32$ & $- 8/3$ & $8$ \\
$\psi^\prime(1/3)$ & $- 1673/243$ & $- 484/243$ & $3064/243$ & $142/27$ & $41081$ \\
$\psi^\prime(1/3) \alpha$ & $109/27$ & $127/54$ & $- 23/27$ & $- 7/18$ & $1/18$ \\
$\psi^\prime(1/3) \alpha^2$ & $- 11/27$ & $5/27/3$ & $13/27$ & $0$ & $- 1/9$ \\
$\psi^\prime(1/3) \alpha^3$ & $0$ & $0$ & $0$ & $0$ & $0$ \\
$\psi^\prime(1/3) \pi^2 $ & $32/81$ & $32/81$ & $0$ & $16/81$ & $32/81$ \\
$(\psi^\prime(1/3))^2$ & $- 8/27$ & $- 8/27$ & $0$ & $- 4/27$ & $- 8/27$ \\
$\psi^{\prime\prime\prime}(1/3)$ & $- 5/18$ & $- 1/9$ & $11/27$ & $7/162$ & $5/162$ \\
$\psi^{\prime\prime\prime}(1/3) \alpha$ & $13/162$ & $1/81$ & $- 4/81$ & $0$ & $1/54$ \\
$\pi^3 \alpha/\sqrt{3}$ & $- 29/81$ & $- 203/1944$ & $29/162$ & $29/1944$ & $- 29/648$ \\
$\pi^3/\sqrt{3}$ & $232/243$ & $725/1944$ & $- 377/243$ & $- 145/648$ & $- 551/1944$ \\
$\pi \ln(3) \alpha/\sqrt{3}$ & $- 4$ & $- 7/6$ & $2$ & $1/6$ & $- 1/2$ \\
$\pi \ln(3)/\sqrt{3}$ & $32/3$ & $25/6$ & $- 52/3$ & $- 5/2$ & $- 19/6$ \\
$\pi (\ln(3))^2 \alpha/\sqrt{3}$ & $1/3$ & $7/72$ & $- 1/6$ & $- 1/72$ & $1/24$ \\
$\pi (\ln(3))^2/\sqrt{3}$ & $- 8/9$ & $- 25/72$ & $13/9$ & $5/24$ & $19/72$ \\
\hline
\end{tabular}
\end{center}
\begin{center}
{Table $8$. Coefficients of $C_F C_A$ for two loop $\MSbar$ $W_2$ amplitudes
continued.}
\end{center}
\end{table}}

\clearpage

{\begin{table}[ht]
\begin{center}
\begin{tabular}{|c||r|r|r|r|r|}
\hline
$a^{(22)}_n$ & $\left. c^{\partial W_2,(22)}_{(1)\,n} \right|_{\MSbars}$ 
& $\left. c^{\partial W_2,(22)}_{(3)\,n} \right|_{\MSbars}$ 
& $\left. c^{\partial W_2,(22)}_{(4)\,n} \right|_{\MSbars}$ 
& $\left. c^{\partial W_2,(22)}_{(5)\,n} \right|_{\MSbars}$ 
& $\left. c^{\partial W_2,(22)}_{(9)\,n} \right|_{\MSbars}$ \\
\hline
$1$ & $169/18$ & $1414/27$ & $707/18$ & $707/27$ & $0$ \\
$\pi^2 \alpha$ & $- 10/27$ & $- 56/27$ & $- 20/27$ & $16/27$ & $2/9$ \\
$\pi^4 \alpha$ & $4/81$ & $8/81$ & $8/81$ & $8/81$ & $- 4/81$ \\
$\zeta(3) \alpha$ & $3$ & $0$ & $0$ & $0$ & $- 2/3$ \\
$\Sigma \alpha$ & $2/3$ & $4/3$ & $4/3$ & $4/3$ & $0$ \\
$\alpha$ & $- 10$ & $- 28/3$ & $- 7$ & $- 14/3$ & $0$ \\
$\pi^2 \alpha^2$ & $- 2/9$ & $- 4/9$ & $- 4/9$ & $- 4/9$ & $2/27$ \\
$\alpha^2$ & $- 15/8$ & $- 2$ & $- 3/2$ & $- 1$ & $0$ \\
$\pi^2 \alpha^3$ & $0$ & $0$ & $0$ & $0$ & $0$ \\
$\alpha^3$ & $0$ & $0$ & $0$ & $0$ & $0$ \\
$\pi^2$ & $-236/243$ & $532/243$ & $- 472/243$ & $- 164/27$ & $- 1672/243$ \\
$\pi^4$ & $- 10/81$ & $- 64/243$ & $- 20/81$ & $- 56/243$ & $- 32/81$ \\
$\zeta(3)$ & $- 2/3$ & $- 4$ & $- 22/3$ & $- 32/3$ & $- 28/3$ \\
$\Sigma$ & $- 2/3$ & $- 8/3$ & $- 4/3$ & $0$ & $- 28/3$ \\
$s_2(\pi/6)$ & $10$ & $0$ & $20$ & $40$ & $68$ \\
$s_2(\pi/6) \alpha$ & $- 6$ & $0$ & $- 12$ & $- 80$ & $4$ \\
$s_2(\pi/2)$ & $86$ & $0$ & $- 40$ & $- 80$ & $- 136$ \\
$s_2(\pi/2) \alpha$ & $12$ & $0$ & $24$ & $48$ & $- 8$ \\
$s_3(\pi/6)$ & $- 50/3$ & $0$ & $- 100/3$ & $- 200/3$ & $- 340/3$ \\
$s_3(\pi/6) \alpha$ & $10$ & $0$ & $20$ & $40$ & $- 20/3$ \\
$s_3(\pi/2)$ & $40/3$ & $0$ & $80/3$ & $160/3$ & $272/3$ \\
$s_3(\pi/2) \alpha$ & $- 8$ & $0$ & $- 16$ & $- 32$ & $16/3$ \\
$\psi^\prime(1/3)$ & $118/81$ & $- 266/81$ & $236/81$ & $82/9$ & $836/81$ \\
$\psi^\prime(1/3) \alpha$ & $5/9$ & $28/9$ & $10/9$ & $- 8/9$ & $- 1/3$ \\
$\psi^\prime(1/3) \alpha^2$ & $1/3$ & $2/3$ & $2/3$ & $2/3$ & $- 1/9$ \\
$\psi^\prime(1/3) \alpha^3$ & $0$ & $0$ & $0$ & $0$ & $0$ \\
$\psi^\prime(1/3) \pi^2 $ & $8/27$ & $64/81$ & $16/27$ & $32/81$ & $16/27$ \\
$(\psi^\prime(1/3))^2$ & $- 2/9$ & $- 16/27$ & $- 4/9$ & $- 8/27$ & $- 4/9$ \\
$\psi^{\prime\prime\prime}(1/3)$ & $1/108$ & $0$ & $1/54$ & $1/27$ & $2/27$ \\
$\psi^{\prime\prime\prime}(1/3) \alpha$ & $- 1/54$ & $- 1/27$ & $- 1/27$ & $- 4/81$ & $1/54$ \\
$\pi^3 \alpha/\sqrt{3}$ & $29/648$ & $0$ & $29/324$ & $29/162$ & $- 29/972$ \\
$\pi^3/\sqrt{3}$ & $- 145/1944$ & $0$ & $- 145/972$ & $- 145/486$ & $- 493/972$ \\
$\pi \ln(3) \alpha/\sqrt{3}$ & $1/2$ & $0$ & $1$ & $2$ & $- 1/3$ \\
$\pi \ln(3)/\sqrt{3}$ & $- 5/6$ & $0$ & $- 5/3$ & $- 10/3$ & $- 17/3$ \\
$\pi (\ln(3))^2 \alpha/\sqrt{3}$ & $- 1/24$ & $0$ & $- 1/12$ & $- 1/6$ & $1/36$ \\
$\pi (\ln(3))^2/\sqrt{3}$ & $5/72$ & $0$ & $5/36$ & $5/18$ & $17/36$ \\
\hline
\end{tabular}
\end{center}
\begin{center}
{Table $9$. Coefficients of $C_F C_A$ for two loop $\MSbar$ $\partial W_2$ 
amplitudes.}
\end{center}
\end{table}}

\clearpage

{\begin{table}[ht]
\begin{center}
\begin{tabular}{|c||r|r|r|r|r|}
\hline
$a^{(23)}_n$ & $\left. c^{W_2,(23)}_{(1)\,n} \right|_{\MSbars}$ 
& $\left. c^{W_2,(23)}_{(2)\,n} \right|_{\MSbars}$ 
& $\left. c^{W_2,(23)}_{(3)\,n} \right|_{\MSbars}$ 
& $\left. c^{W_2,(23)}_{(4)\,n} \right|_{\MSbars}$ 
& $\left. c^{W_2,(23)}_{(5)\,n} \right|_{\MSbars}$ \\
\hline
$1$ & $3971/324$ & $- 9805/648$ & $1238/81$ & $- 1669/162$ & $- 1987/81$ \\
$\pi^2 \alpha$ & $248/243$ & $- 824/243$ & $1288/729$ & $- 2296/729$ & $- 6152/729$ \\
$\pi^4 \alpha$ & $0$ & $16/81$ & $32/243$ & $64/243$ & $128/243$ \\
$\zeta(3) \alpha$ & $- 4/3$ & $8/3$ & $- 8/3$ & $8/3$ & $8$ \\
$\Sigma \alpha$ & $0$ & $4/3$ & $8/9$ & $16/9$ & $32/9$ \\
$\alpha$ & $- 107/108$ & $539/108$ & $32/81$ & $- 1697/162$ & $1838/81$ \\
$\pi^2 \alpha^2$ & $- 2/27$ & $- 2/9$ & $- 20/81$ & $- 28/81$ & $- 44/81$ \\
$\pi^4 \alpha^2$ & $0$ & $0$ & $0$ & $0$ & $0$ \\
$\alpha^2$ & $- 5/12$ & $- 7/12$ & $- 10/9$ & $- 25/18$ & $- 16/9$ \\
$\pi^2$ & $- 869/243$ & $1967/243$ & $- 8852/729$ & $9362/729$ & $15424/729$ \\
$\pi^4$ & $- 68/243$ & $8/243$ & $- 832/729$ & $280/729$ & $608/729$ \\
$\zeta(3)$ & $- 46/3$ & $6$ & $- 136/3$ & $12$ & $32$ \\
$\Sigma$ & $4/9$ & $- 10/9$ & $8/27$ & $- 32/27$ & $- 40/27$ \\
$s_2(\pi/6)$ & $104$ & $- 96$ & $320$ & $- 144$ & $- 288$ \\
$s_2(\pi/6) \alpha$ & $- 16$ & $32$ & $- 32$ & $32$ & $96$ \\
$s_2(\pi/2)$ & $- 208$ & $192$ & $- 640$ & $288$ & $576$ \\
$s_2(\pi/2) \alpha$ & $32$ & $- 64$ & $64$ & $- 64$ & $- 192$ \\
$s_3(\pi/6)$ & $- 520/3$ & $160$ & $- 1600/3$ & $240$ & $480$ \\
$s_3(\pi/6) \alpha$ & $80/3$ & $- 160/3$ & $- 1600/3$ & $- 160/3$ & $- 160$ \\
$s_3(\pi/2)$ & $416/3$ & $- 128$ & $1280/3$ & $- 192$ & $- 384$ \\
$s_3(\pi/2) \alpha$ & $- 64/3$ & $128/3$ & $- 128/3$ & $128/3$ & $128$ \\
$\psi^\prime(1/3)$ & $869/162$ & $- 1967/162$ & $4426/243$ & $- 4681/243$ & $- 7712/243$ \\
$\psi^\prime(1/3) \pi^2 \alpha$ & $0$ & $0$ & $0$ & $0$ & $0$ \\
$\psi^\prime(1/3) \alpha$ & $- 124/81$ & $412/81$ & $- 644/243$ & $1148/243$ & $3076/243$ \\
$\psi^\prime(1/3) \pi^2 \alpha^2 $ & $0$ & $0$ & $0$ & $0$ & $0$ \\
$\psi^\prime(1/3) \alpha^2 $ & $1/9$ & $1/3$ & $10/27$ & $14/27$ & $22/27$ \\
$\psi^\prime(1/3) \pi^2 $ & $- 16/27$ & $0$ & $- 128/81$ & $- 32/81$ & $- 76/243$ \\
$(\psi^\prime(1/3))^2$ & $4/9$ & $0$ & $32/27$ & $8/27$ & $0$ \\
$(\psi^\prime(1/3))^2 \alpha$ & $0$ & $0$ & $0$ & $0$ & $0$ \\
$(\psi^\prime(1/3))^2 \alpha^2$ & $0$ & $0$ & $0$ & $0$ & $0$ \\
$\psi^{\prime\prime\prime}(1/3)$ & $29/162$ & $- 1/81$ & $- 23/243$ & $- 76/243$ & $- 76/243$ \\
$\psi^{\prime\prime\prime}(1/3) \alpha$ & $0$ & $- 2/27$ & $- 4/81$ & $- 8/81$ & $- 16/81$ \\
$\pi^3 \alpha/\sqrt{3}$ & $29/162$ & $- 58/243$ & $58/243$ & $- 58/243$ & $- 58/81$ \\
$\pi^3/\sqrt{3}$ & $- 377/486$ & $58/81$ & $- 580/243$ & $29/27$ & $- 58/27$ \\
$\pi \ln(3) \alpha/\sqrt{3}$ & $4/3$ & $- 8/3$ & $8/3$ & $- 8/3$ & $- 8$ \\
$\pi \ln(3)/\sqrt{3}$ & $- 26/3$ & $8$ & $- 80/3$ & $12$ & $24$ \\
$\pi (\ln(3))^2 \alpha/\sqrt{3}$ & $- 1/9$ & $2/9$ & $- 2/9$ & $2/9$ & $2/3$ \\
$\pi (\ln(3))^2/\sqrt{3}$ & $13/18$ & $- 2/3$ & $20/9$ & $- 1$ & $- 2$ \\
\hline
\end{tabular}
\end{center}
\begin{center}
{Table $10$. Coefficients of $C_F^2$ for two loop $\MSbar$ $W_2$ amplitudes.}
\end{center}
\end{table}}

\clearpage 

{\begin{table}[hb]
\begin{center}
\begin{tabular}{|c||r|r|r|r|r|}
\hline
$a^{(23)}_n$ & $\left. c^{W_2,(23)}_{(6)\,n} \right|_{\MSbars}$ 
& $\left. c^{W_2,(23)}_{(7)\,n} \right|_{\MSbars}$ 
& $\left. c^{W_2,(23)}_{(8)\,n} \right|_{\MSbars}$ 
& $\left. c^{W_2,(23)}_{(9)\,n} \right|_{\MSbars}$ 
& $\left. c^{W_2,(23)}_{(10)\,n} \right|_{\MSbars}$ \\
\hline
$1$ & $1609/81$ & $535/162$ & $- 1994/81$ & $185/54$ & $- 185/54$ \\
$\pi^2 \alpha$ & $320/729$ & $- 1160/729$ & $- 2368/729$ & $0$ & $0$ \\
$\pi^4 \alpha$ & $- 32/243$ & $32/243$ & $64/243$ & $0$ & $0$ \\
$\zeta(3) \alpha$ & $- 8/3$ & $0$ & $8/3$ & $0$ & $0$ \\
$\Sigma \alpha$ & $- 8/9$ & $8/9$ & $16/9$ & $0$ & $0$ \\
$\alpha$ & $- 1406/81$ & $- 401/162$ & $832/81$ & $- 5/6$ & $5/6$ \\
$\pi^2 \alpha^2$ & $- 4/81$ & $- 20/81$ & $- 28/81$ & $0$ & $0$ \\
$\pi^4 \alpha^2$ & $0$ & $0$ & $0$ & $0$ & $0$ \\
$\alpha^2$ & $4/9$ & $- 11/18$ & $- 14/9$ & $1/6$ & $- 1/6$ \\
$\pi^2$ & $- 7000/729$ & $- 2774/729$ & $13604/729$ & $1858/243$ & $2786/243$ \\
$\pi^4$ & $- 560/729$ & $- 640/729$ & $64/729$ & $104/243$ & $- 32/243$ \\
$\zeta(3)$ & $- 112/3$ & $- 92/3$ & $40/3$ & $28/2$ & $- 4/3$ \\
$\Sigma$ & $40/27$ & $- 4/27$ & $- 80/27$ & $0$ & $- 4/3$  \\
$s_2(\pi/6)$ & $224$ & $160$ & $- 224$ & $- 80$ & $- 64$ \\
$s_2(\pi/6) \alpha$ & $- 32$ & $0$ & $32$ & $0$ & $0$ \\
$s_2(\pi/2)$ & $- 448$ & $- 320$ & $448$ & $160$ & $128$ \\
$s_2(\pi/2) \alpha$ & $64$ & $0$ & $- 64$ & $0$ & $0$ \\
$s_3(\pi/6)$ & $- 1120/3$ & $- 800/3$ & $1120/3$ & $- 320/3$ & $320/3$ \\
$s_3(\pi/6) \alpha$ & $160/3$ & $0$ & $- 160/3$ & $0$ & $0$ \\
$s_3(\pi/2)$ & $896/3$ & $640/3$ & $- 896/3$ & $- 320/3$ & $- 256/3$ \\
$s_3(\pi/2) \alpha$ & $- 128/3$ & $0$ & $128/3$ & $0$ & $0$ \\
$\psi^\prime(1/3)$ & $3500/243$ & $1387/243$ & $- 6802/243$ & $- 929/81$ & $- 1393/81$ \\
$\psi^\prime(1/3) \pi^2 \alpha$ & $0$ & $0$ & $0$ & $0$ & $0$ \\
$\psi^\prime(1/3) \alpha$ & $- 160/243$ & $580/243$ & $1184/243$ & $0$ & $0$ \\
$\psi^\prime(1/3) \pi^2 \alpha^2 $ & $0$ & $0$ & $0$ & $0$ & $0$ \\
$\psi^\prime(1/3) \alpha^2 $ & $2/27$ & $10/27$ & $14/27$ & $0$ & $2/9$ \\
$\psi^\prime(1/3) \pi^2 $ & $- 64/81$ & $- 64/81$ & $0$ & $- 32/81$ & $- 64/81$ \\
$(\psi^\prime(1/3))^2$ & $16/27$ & $16/27$ & $0$ & $8/27$ & $16/27$ \\
$(\psi^\prime(1/3))^2 \alpha$ & $0$ & $0$ & $0$ & $0$ & $0$ \\
$(\psi^\prime(1/3))^2 \alpha^2$ & $0$ & $0$ & $0$ & $0$ & $0$ \\
$\psi^{\prime\prime\prime}(1/3)$ & $94/243$ & $104/243$ & $- 8/243$ & $- 1/9$ & $4/27$ \\
$\psi^{\prime\prime\prime}(1/3) \alpha$ & $4/81$ & $- 4/81$ & $- 8/81$ & $0$ & $0$ \\
$\pi^3 \alpha/\sqrt{3}$ & $58/243$ & $0$ & $- 58/243$ & $0$ & $0$ \\
$\pi^3/\sqrt{3}$ & $- 406/243$ & $- 290/243$ & $406/243$ & $145/243$ & $116/243$ \\
$\pi \ln(3) \alpha/\sqrt{3}$ & $8/3$ & $0$ & $- 8/3$ & $0$ & $0$ \\
$\pi \ln(3)/\sqrt{3}$ & $- 56/3$ & $- 40/3$ & $56/3$ & $20/3$ & $16/3$ \\
$\pi (\ln(3))^2 \alpha/\sqrt{3}$ & $- 2/9$ & $0$ & $2/9$ & $0$ & $0$ \\
$\pi (\ln(3))^2/\sqrt{3}$ & $14/9$ & $10/9$ & $- 14/9$ & $- 5/9$ & $- 4/9$ \\
\hline
\end{tabular}
\end{center}
\begin{center}
{Table $11$. Coefficients of $C_F^2$ for two loop $\MSbar$ $W_2$ amplitudes 
continued.}
\end{center}
\end{table}}

\clearpage

{\begin{table}[ht]
\begin{center}
\begin{tabular}{|c||r|r|r|r|r|}
\hline
$a^{(23)}_n$ & $\left. c^{\partial W_2,(23)}_{(1)\,n} \right|_{\MSbars}$ 
& $\left. c^{\partial W_2,(23)}_{(3)\,n} \right|_{\MSbars}$ 
& $\left. c^{\partial W_2,(23)}_{(4)\,n} \right|_{\MSbars}$ 
& $\left. c^{\partial W_2,(23)}_{(5)\,n} \right|_{\MSbars}$ 
& $\left. c^{\partial W_2,(23)}_{(9)\,n} \right|_{\MSbars}$ \\
\hline
$1$ & $- 23/8$ & $- 28/3$ & $- 7$ & $- 14/3$ & $0$ \\
$\pi^2 \alpha$ & $- 64/27$ & $- 40/27$ & $- 128/27$ & $- 8$ & $- 4/27$ \\
$\pi^4 \alpha$ & $16/81$ & $32/81$ & $32/81$ & $32/81$ & $0$ \\
$\zeta(3) \alpha$ & $4/3$ & $0$ & $8/3$ & $16/3$ & $0$ \\
$\Sigma \alpha$ & $4/3$ & $8/3$ & $8/3$ & $8/3$ & $0$ \\
$\alpha$ & $4$ & $32/3$ & $32/81$ & $16/3$ & $0$ \\
$\pi^2 \alpha^2$ & $- 8/27$ & $- 16/27$ & $- 16/27$ & $- 16/27$ & $0$ \\
$\pi^4 \alpha^2$ & $0$ & $0$ & $0$ & $0$ & $0$ \\
$\alpha^2$ & $- 1$ & $- 8/3$ & $- 2$ & $- 4/3$ & $0$ \\
$\pi^2$ & $122/27$ & $176/27$ & $244/27$ & $104/9$ & $172/9$ \\
$\pi^4$ & $- 20/81$ & $- 256/243$ & $- 40/81$ & $16/243$ & $8/27$ \\
$\zeta(3)$ & $- 28/3$ & $- 32$ & $- 56/3$ & $- 16/3$ & $8$ \\
$\Sigma$ & $- 2/3$ & $- 8/3$ & $- 4/3$ & $0$ & $- 4/3$ \\
$s_2(\pi/6)$ & $8$ & $96$ & $16$ & $- 64$ & $- 144$ \\
$s_2(\pi/6) \alpha$ & $16$ & $0$ & $32$ & $64$ & $0$ \\
$s_2(\pi/2)$ & $- 16$ & $- 192$ & $- 64$ & $128$ & $288$ \\
$s_2(\pi/2) \alpha$ & $- 32$ & $0$ & $- 80/3$ & $- 128$ & $0$ \\
$s_3(\pi/6)$ & $- 40/3$ & $-160$ & $- 160/3$ & $3203$ & $240$ \\
$s_3(\pi/6) \alpha$ & $- 80/3$ & $0$ & $64/3$ & $- 320/3$ & $0$ \\
$s_3(\pi/2)$ & $32/3$ & $128$ & $128/3$ & $- 256/3$ & $- 192$ \\
$s_3(\pi/2) \alpha$ & $64/3$ & $0$ & $- 121/3$ & $256/3$ & $0$ \\
$\psi^\prime(1/3)$ & $- 61/9$ & $- 88/9$ & $- 122/9$ & $- 52/3$ & $- 86/3$ \\
$\psi^\prime(1/3) \pi^2 \alpha$ & $0$ & $0$ & $0$ & $0$ & $0$ \\
$\psi^\prime(1/3) \alpha$ & $32/9$ & $20/9$ & $64/9$ & $12$ & $0$ \\
$\psi^\prime(1/3) \pi^2 \alpha^2 $ & $0$ & $0$ & $0$ & $0$ & $0$ \\
$\psi^\prime(1/3) \alpha^2 $ & $4/9$ & $8/9$ & $8/9$ & $8/9$ & $2/9$ \\
$\psi^\prime(1/3) \pi^2 $ & $- 16/27$ & $- 32/27$ & $- 128/81$ & $- 64/81$ & $- 32/27$ \\
$(\psi^\prime(1/3))^2$ & $4/9$ & $32/27$ & $8/9$ & $16/27$ & $8/9$ \\
$(\psi^\prime(1/3))^2 \alpha$ & $0$ & $0$ & $0$ & $0$ & $0$ \\
$(\psi^\prime(1/3))^2 \alpha^2$ & $0$ & $0$ & $0$ & $0$ & $0$ \\
$\psi^{\prime\prime\prime}(1/3)$ & $1/6$ & $32/27$ & $1/3$ & $2/7$ & $1/27$ \\
$\psi^{\prime\prime\prime}(1/3) \alpha$ & $- 2/27$ & $- 4/27$ & $- 4/27$ & $- 8/81$ & $0$ \\
$\pi^3 \alpha/\sqrt{3}$ & $- 29/243$ & $0$ & $- 58/243$ & $- 116/243$ & $0$ \\
$\pi^3/\sqrt{3}$ & $- 29/486$ & $- 58/81$ & $- 29/243$ & $116/243$ & $29/27$ \\
$\pi \ln(3) \alpha/\sqrt{3}$ & $- 4/3$ & $0$ & $- 8/3$ & $- 16/3$ & $0$ \\
$\pi \ln(3)/\sqrt{3}$ & $- 2/3$ & $- 8$ & $- 4/3$ & $16/3$ & $12$ \\
$\pi (\ln(3))^2 \alpha/\sqrt{3}$ & $1/9$ & $0$ & $2/9$ & $4/9$ & $0$ \\
$\pi (\ln(3))^2/\sqrt{3}$ & $1/18$ & $2/3$ & $1/9$ & $- 4/9$ & $- 1$ \\
\hline
\end{tabular}
\end{center}
\begin{center}
{Table $12$. Coefficients of $C_F^2$ for two loop $\MSbar$ $\partial W_2$ 
amplitudes.}
\end{center}
\end{table}}

\clearpage 

{\begin{table}[ht]
\begin{center}
\begin{tabular}{|c||r|r|r|r|r|r|r|r|}
\hline
$a^{(1)}_n$ & $c^{W_2,(1)}_{(3)\,n}$ & $c^{W_2,(1)}_{(4)\,n}$ 
& $c^{W_2,(1)}_{(5)\,n}$ & $c^{W_2,(1)}_{(6)\,n}$ 
& $c^{W_2,(1)}_{(7)\,n}$ & $c^{W_2,(1)}_{(8)\,n}$ 
& $c^{W_2,(1)}_{(9)\,n}$ & $c^{W_2,(1)}_{(10)\,n}$ \\
\hline
$1$ & $16/27$ & $71/27$ & $100/27$ & $- 28/27$ & $37/27$ & $128/27$ & $- 1/3$ & $1/3$ \\
$\pi^2 \alpha$ & $- 16/81$ & $- 32/81$ & $- 64/81$ & $16/81$ & $- 16/81$ & $- 32/81$ & $0$ & $0$ \\
$\alpha$ & $- 8/9$ & $- 16/9$ & $- 26/9$ & $14/9$ & $- 2/9$ & $- 16/9$ & $0$ & $0$ \\
$\pi^2$ & $- 16/243$ & $64/243$ & $80/243$ & $- 80/243$ & $8/243$ & $160/243$ & $0$ & $8/27$ \\
$\psi^\prime(1/3)$ & $8/81$ & $- 32/81$ & $- 40/81$ & $40/81$ & $- 4/81$ & $- 80/81$ & $0$ & $0$ \\
$\psi^\prime(1/3) \alpha$ & $8/27$ & $16/27$ & $40/27$ & $- 8/27$ & $8/27$ & $16/27$ & $0$ & $- 4/9$ \\
\hline
\end{tabular}
\end{center}
\begin{center}
{Table $13$. Coefficients of $C_F$ for one loop RI${}^\prime$/SMOM $W_2$ 
amplitudes.}
\end{center}
\end{table}}

\vspace{3cm}
\noindent
{\begin{table}[ht]
\begin{center}
\begin{tabular}{|c||r|r|r|r|r|}
\hline
$a^{(1)}_n$ & $c^{\partial W_2,(1)}_{(1)\,n}$ & $c^{\partial W_2,(1)}_{(3)\,n}$ 
& $c^{\partial W_2,(1)}_{(4)\,n}$ & $c^{\partial W_2,(1)}_{(5)\,n}$ 
& $c^{\partial W_2,(1)}_{(9)\,n}$ \\
\hline
$1$ & $2$ & $16/3$ & $4$ & $8/3$ & $0$ \\
$\pi^2 \alpha$ & $- 8/27$ & $- 16/27$ & $- 16/27$ & $- 16/27$ & $0$ \\
$\alpha$ & $- 1$ & $- 8/3$ & $- 2$ & $- 4/3$ & $0$ \\
$\pi^2$ & $4/27$ & $16/27$ & $8/27$ & $0$ & $8/27$ \\
$\psi^\prime(1/3)$ & $- 2/9$ & $- 8/9$ & $- 4/9$ & $0$ & $- 4/9$ \\
$\psi^\prime(1/3) \alpha$ & $4/9$ & $8/9$ & $8/9$ & $8/9$ & $0$ \\
\hline
\end{tabular}
\end{center}
\begin{center}
{Table $14$. Coefficients of $C_F$ for one loop RI${}^\prime$/SMOM 
$\partial W_2$ amplitudes.}
\end{center}
\end{table}}

\vspace{3cm}
{\begin{table}[hb]
\begin{center}
\begin{tabular}{|c||r|r|r|r|}
\hline
$a^{(21)}_n$ & $c^{W_2,(21)}_{(3)\,n}$ & $c^{W_2,(21)}_{(4)\,n}$ 
& $c^{W_2,(21)}_{(5)\,n}$ & $c^{W_2,(21)}_{(6)\,n}$ \\ 
\hline
$1$ & $8/27$ & $- 8$ & $- 308/27$ & $76/27$ \\
$\pi^2 \alpha$ & $320/729$ & $640/729$ & $1280/729$ & $- 320/729$ \\
$\alpha$ & $160/81$ & $320/81$ & $520/81$ & $- 280/81$ \\
$\pi^2$ & $80/243$ & $- 248/243$ & $- 64/81$ & $64/81$ \\
$\psi^\prime(1/3)$ & $- 40/81$ & $124/81$ & $32/27$ & $- 32/27$ \\
$\psi^\prime(1/3) \alpha$ & $- 160/243$ & $- 320/243$ & $- 640/243$ & $160/243$ \\
\hline
\end{tabular}
\end{center}
\begin{center}
{Table $15$. Coefficients of $C_F T_F \Nf$ for two loop RI${}^\prime$/SMOM 
$W_2$ amplitudes.}
\end{center}
\end{table}}

\clearpage 

{\begin{table}[hb]
\begin{center}
\begin{tabular}{|c||r|r|r|r|}
\hline
$a^{(21)}_n$ & $c^{W_2,(21)}_{(7)\,n}$ & $c^{W_2,(21)}_{(8)\,n}$ 
& $c^{W_2,(21)}_{(9)\,n}$ & $c^{W_2,(21)}_{(10)\,n}$ \\
\hline
$1$ & $- 44/9$ & $- 472/27$ & $32/27$ & $- 32/27$ \\
$\pi^2 \alpha$ & $320/729$ & $640/729$ & $0$ & $0$ \\
$\alpha$ & $40/81$ & $320/81$ & $0$ & $0$ \\
$\pi^2$ & $- 56/243$ & $- 688/243$ & $8/81$ & $- 232/243$ \\
$\psi^\prime(1/3)$ & $28/81$ & $344/81$ & $- 4/27$ & $116/81$ \\
$\psi^\prime(1/3) \alpha$ & $- 160/243$ & $- 320/243$ & $0$ & $0$ \\
\hline
\end{tabular}
\end{center}
\begin{center}
{Table $16$. Coefficients of $C_F T_F \Nf$ for two loop RI${}^\prime$/SMOM 
$W_2$ amplitudes continued.}
\end{center}
\end{table}}

\vspace{5cm}
{\begin{table}[hb]
\begin{center}
\begin{tabular}{|c||r|r|r|r|r|}
\hline
$a^{(21)}_n$ & $c^{\partial W_2,(21)}_{(1)\,n}$ & 
$c^{\partial W_2,(21)}_{(3)\,n}$ & $c^{\partial W_2,(21)}_{(4)\,n}$ & 
$c^{\partial W_2,(21)}_{(5)\,n}$ & $c^{\partial W_2,(21)}_{(9)\,n}$ \\ 
\hline
$1$ & $- 58/9$ & $- 464/27$ & $- 116/9$ & $- 232/27$ & $0$ \\
$\pi^2 \alpha$ & $160/243$ & $320/243$ & $320/243$ & $320/243$ & $0$ \\
$\alpha$ & $20/9$ & $160/27$ & $40/9$ & $80/27$ & $0$ \\
$\pi^2$ & $- 152/243$ & $- 608/243$ & $- 304/243$ & $0$ & $- 208/243$ \\
$\psi^\prime(1/3)$ & $76/81$ & $304/81$ & $152/81$ & $0$ & $104/81$ \\
$\psi^\prime(1/3) \alpha$ & $- 80/81$ & $- 160/81$ & $- 160/81$ & $- 160/81$ & $0$ \\
\hline
\end{tabular}
\end{center}
\begin{center}
{Table $17$. Coefficients of $C_F T_F \Nf$ for two loop RI${}^\prime$/SMOM 
$\partial W_2$ amplitudes.}
\end{center}
\end{table}}

\clearpage

{\begin{table}[ht]
\begin{center}
\begin{tabular}{|c||r|r|r|r|}
\hline
$a^{(22)}_n$ & $c^{W_2,(22)}_{(3)\,n}$ & $c^{W_2,(22)}_{(4)\,n}$ 
& $c^{W_2,(22)}_{(5)\,n}$ & $c^{W_2,(22)}_{(6)\,n}$ \\
\hline
$1$ & $- 188/81$ & $8351/324$ & $5879/162$ & $- 1637/162$ \\
$\pi^2 \alpha$ & $- 2314/729$ & $- 173/729$ & $842/729$ & $- 1574/729$ \\
$\pi^4 \alpha$ & $- 8/243$ & $32/243$ & $76/243$ & $- 52/243$ \\
$\zeta(3) \alpha$ & $- 2$ & $7/3$ & $6$ & $- 6$ \\
$\Sigma \alpha$ & $4/9$ & $8/9$ & $16/9$ & $- 4/9$ \\
$\alpha$ & $- 401/81$ & $- 1901/162$ & $- 3187/162$ & $1849/162$ \\
$\pi^2 \alpha^2$ & $- 2/9$ & $- 14/27$ & $- 10/9$ & $10/27$ \\
$\alpha^2$ & $- 1$ & $- 29/12$ & $- 25/6$ & $5/2$ \\
$\pi^2 \alpha^3$ & $- 4/81$ & $- 8/81$ & $- 16/81$ & $4/81$ \\
$\alpha^3$ & $- 2/9$ & $- 4/9$ & $- 13/18$ & $7/18$ \\
$\pi^2$ & $7724/729$ & $- 2384/729$ & $- 7774/729$ & $3346/729$ \\
$\pi^4$ & $200/243$ & $- 100/243$ & $- 68/81$ & $148/243$ \\
$\zeta(3)$ & $98/3$ & $- 41/3$ & $- 34$ & $70/3$ \\
$\Sigma$ & $4/9$ & $- 10/9$ & $- 8/9$ & $8/9$ \\
$s_2(\pi/6)$ & $- 208$ & $70$ & $168$ & $- 128$ \\
$s_2(\pi/6) \alpha$ & $24$ & $- 26$ & $- 72$ & $48$ \\
$s_2(\pi/2)$ & $416$ & $- 140$ & $- 336$ & $256$ \\
$s_2(\pi/2) \alpha$ & $- 48$ & $52$ & $144$ & $- 96$ \\
$s_3(\pi/6)$ & $1040/3$ & $- 350/3$ & $- 280$ & $640/3$ \\
$s_3(\pi/6) \alpha$ & $- 40$ & $130/3$ & $120$ & $-80$ \\
$s_3(\pi/2)$ & $- 832/3$ & $280/3$ & $224$ & $- 512/3$ \\
$s_3(\pi/2) \alpha$ & $32$ & $- 104/3$ & $- 96$ & $64$ \\
$\psi^\prime(1/3)$ & $- 3862/243$ & $1192/243$ & $3887/243$ & $- 1673/243$ \\
$\psi^\prime(1/3) \alpha$ & $1157/243$ & $173/486$ & $- 421/243$ & $787/243$ \\
$\psi^\prime(1/3) \alpha^2$ & $1/3$ & $7/9$ & $5/3$ & $- 5/9$ \\
$\psi^\prime(1/3) \alpha^3$ & $2/27$ & $4/27$ & $8/27$ & $- 2/27$ \\
$\psi^\prime(1/3) \pi^2 $ & $64/81$ & $16/81$ & $17/54$ & $32/81$ \\
$(\psi^\prime(1/3))^2$ & $- 16/27$ & $- 4/27$ & $- 19/162$ & $- 8/27$ \\
$\psi^{\prime\prime\prime}(1/3)$ & $- 11/27$ & $7/54$ & $17/54$ & $- 5/18$ \\
$\psi^{\prime\prime\prime}(1/3) \alpha$ & $1/81$ & $- 4/81$ & $- 19/162$ & $13/162$ \\
$\pi^3 \alpha/\sqrt{3}$ & $- 29/162$ & $377/1944$ & $29/54$ & $- 29/81$ \\
$\pi^3/\sqrt{3}$ & $377/243$ & $1015/1944$ & $- 203/162$ & $232/243$ \\
$\pi \ln(3) \alpha/\sqrt{3}$ & $- 2$ & $13/6$ & $6$ & $- 4$ \\
$\pi \ln(3)/\sqrt{3}$ & $52/3$ & $- 35/6$ & $- 14$ & $32/3$ \\
$\pi (\ln(3))^2 \alpha/\sqrt{3}$ & $1/6$ & $- 13/72$ & $- 1/2$ & $1/3$ \\
$\pi (\ln(3))^2/\sqrt{3}$ & $- 13/9$ & $35/72$ & $7/6$ & $- 8/9$ \\
\hline
\end{tabular}
\end{center}
\begin{center}
{Table $18$. Coefficients of $C_F C_A$ for two loop RI${}^\prime$/SMOM $W_2$ 
amplitudes.}
\end{center}
\end{table}}

\clearpage

{\begin{table}[ht]
\begin{center}
\begin{tabular}{|c||r|r|r|r|}
\hline
$a^{(22)}_n$ & $c^{W_2,(22)}_{(7)\,n}$ & $c^{W_2,(22)}_{(8)\,n}$ 
& $c^{W_2,(22)}_{(9)\,n}$ & $c^{W_2,(22)}_{(10)\,n}$ \\
\hline
$1$ & $4375/324$ & $4430/81$ & $- 379/108$ & $379/108$ \\
$\pi^2 \alpha$ & $- 1531/729$ & $- 362/729$ & $7/27$ & $- 1/27$ \\
$\pi^4 \alpha$ & $- 8/243$ & $32/243$ & $0$ & $- 4/81$ \\
$\zeta(3) \alpha$ & $- 7/3$ & $2$ & $1/3$ & $- 1$ \\
$\Sigma \alpha$ & $4/9$ & $8/9$ & $0$ & $0$ \\
$\alpha$ & $- 53/81$ & $- 937/81$ & $- 1/6$ & $1/6$ \\
$\pi^2 \alpha^2$ & $- 2/9$ & $- 14/27$ & $0$ & $2/27$ \\
$\alpha^2$ & $- 1/12$ & $- 7/3$ & $- 1/12$ & $1/12$ \\
$\pi^2 \alpha^3$ & $- 4/81$ & $- 8/81$ & $0$ & $0$ \\
$\alpha^3$ & $- 1/18$ & $- 4/9$ & $0$ & $0$ \\
$\pi^2$ & $968/729$ & $- 6128/729$ & $- 284/81$ & $- 820/243$ \\
$\pi^4$ & $40/243$ & $- 88/81$ & $- 44/243$ & $- 52/243$ \\
$\zeta(3)$ & $19/3$ & $- 110/3$ & $- 5$ & $- 13/3$ \\
$\Sigma$ & $- 2/9$ & $- 28/9$ & $- 2/9$ & $- 10/9$ \\
$s_2(\pi/6)$ & $- 50$ & $208$ & $30$ & $38$ \\
$s_2(\pi/6) \alpha$ & $14$ & $- 24$ & $-2$ & $6$ \\
$s_2(\pi/2)$ & $100$ & $- 416$ & $- 60$ & $- 76$ \\
$s_2(\pi/2) \alpha$ & $- 28$ & $48$ & $4$ & $- 12$ \\
$s_3(\pi/6)$ & $250/3$ & $- 1040/3$ & $- 50$ & $- 190/3$ \\
$s_3(\pi/6) \alpha$ & $- 70/3$ & $40$ & $- 50$ & $- 10$ \\
$s_3(\pi/2)$ & $- 200/3$ & $832/3$ & $10/3$ & $152/3$ \\
$s_3(\pi/2) \alpha$ & $56/3$ & $- 32$ & $40$ & $8$ \\
$\psi^\prime(1/3)$ & $- 484/243$ & $3064/243$ & $142/27$ & $410/82$ \\
$\psi^\prime(1/3) \alpha$ & $1531/486$ & $181/243$ & $- 7/18$ & $1/18$ \\
$\psi^\prime(1/3) \alpha^2$ & $1/3$ & $7/9$ & $0$ & $- 1/9$ \\
$\psi^\prime(1/3) \alpha^3$ & $2/27$ & $4/27$ & $0$ & $0$ \\
$\psi^\prime(1/3) \pi^2 $ & $32/81$ & $0$ & $16/81$ & $32/81$ \\
$(\psi^\prime(1/3))^2$ & $- 8/27$ & $0$ & $- 4/27$ & $- 8/27$ \\
$\psi^{\prime\prime\prime}(1/3)$ & $- 1/9$ & $11/27$ & $7/162$ & $5/162$ \\
$\psi^{\prime\prime\prime}(1/3) \alpha$ & $1/81$ & $- 4/81$ & $0$ & $1/54$ \\
$\pi^3 \alpha/\sqrt{3}$ & $- 203/1944$ & $29/162$ & $29/1944$ & $- 29/648$ \\
$\pi^3/\sqrt{3}$ & $725/1944$ & $- 377/243$ & $- 145/648$ & $- 551/648$ \\
$\pi \ln(3) \alpha/\sqrt{3}$ & $- 7/6$ & $2$ & $1/6$ & $- 1/2$ \\
$\pi \ln(3)/\sqrt{3}$ & $25/6$ & $- 52/3$ & $- 5/2$ & $- 19/6$ \\
$\pi (\ln(3))^2 \alpha/\sqrt{3}$ & $7/72$ & $- 1/6$ & $- 1/72$ & $1/24$ \\
$\pi (\ln(3))^2/\sqrt{3}$ & $- 25/72$ & $13/9$ & $5/24$ & $19/72$ \\
\hline
\end{tabular}
\end{center}
\begin{center}
{Table $19$. Coefficients of $C_F C_A$ for two loop RI${}^\prime$/SMOM $W_2$ 
amplitudes
continued.}
\end{center}
\end{table}}

\clearpage

{\begin{table}[ht]
\begin{center}
\begin{tabular}{|c||r|r|r|r|r|}
\hline
$a^{(22)}_n$ & $c^{\partial W_2,(22)}_{(1)\,n}$ & 
$c^{\partial W_2,(22)}_{(3)\,n}$ & $c^{\partial W_2,(22)}_{(4)\,n}$ & 
$c^{\partial W_2,(22)}_{(5)\,n}$ & $c^{\partial W_2,(22)}_{(9)\,n}$ \\
\hline
$1$ & $707/36$ & $1414/27$ & $707/18$ & $707/27$ & $0$ \\
$\pi^2 \alpha$ & $- 284/243$ & $- 892/243$ & $- 568/243$ & $- 244/243$ & $2/9$ \\
$\pi^4 \alpha$ & $4/81$ & $8/81$ & $8/81$ & $8/81$ & $- 4/81$ \\
$\zeta(3) \alpha$ & $0$ & $0$ & $0$ & $0$ & $- 2/3$ \\
$\Sigma \alpha$ & $2/3$ & $4/3$ & $4/3$ & $4/3$ & $0$ \\
$\alpha$ & $- 223/36$ & $- 446/27$ & $- 223/18$ & $- 223/27$ & $0$ \\
$\pi^2 \alpha^2$ & $- 10/27$ & $- 20/27$ & $- 20/27$ & $- 20/27$ & $2/27$ \\
$\alpha^2$ & $- 5/4$ & $- 10/3$ & $- 5/2$ & $- 5/3$ & $0$ \\
$\pi^2 \alpha^3$ & $- 2/27$ & $- 4/27$ & $- 4/27$ & $- 4/27$ & $0$ \\
$\alpha^3$ & $- 1/4$ & $- 2/3$ & $- 1/2$ & $- 1/3$ & $0$ \\
$\pi^2$ & $- 236/243$ & $532/243$ & $- 472/243$ & $- 164/27$ & $- 1672/243$ \\
$\pi^4$ & $- 10/81$ & $- 64/243$ & $- 20/81$ & $- 56/243$ & $- 32/81$ \\
$\zeta(3)$ & $- 11/3$ & $- 4$ & $- 22/3$ & $- 32/3$ & $- 28/3$ \\
$\Sigma$ & $- 2/3$ & $- 8/3$ & $- 4/3$ & $0$ & $- 4/3$ \\
$s_2(\pi/6)$ & $10$ & $0$ & $20$ & $40$ & $68$ \\
$s_2(\pi/6) \alpha$ & $- 6$ & $0$ & $- 12$ & $- 24$ & $4$ \\
$s_2(\pi/2)$ & $- 20$ & $0$ & $- 40$ & $- 80$ & $- 136$ \\
$s_2(\pi/2) \alpha$ & $12$ & $0$ & $24$ & $48$ & $- 8$ \\
$s_3(\pi/6)$ & $- 50/3$ & $0$ & $- 100/3$ & $- 200/3$ & $- 340/3$ \\
$s_3(\pi/6) \alpha$ & $10$ & $0$ & $20$ & $40$ & $- 20/3$ \\
$s_3(\pi/2)$ & $40/3$ & $0$ & $80/3$ & $160/3$ & $272/3$ \\
$s_3(\pi/2) \alpha$ & $- 8$ & $0$ & $- 16$ & $- 32$ & $16/3$ \\
$\psi^\prime(1/3)$ & $118/81$ & $- 266/81$ & $236/81$ & $82/9$ & $836/81$ \\
$\psi^\prime(1/3) \alpha$ & $142/81$ & $446/81$ & $284/81$ & $122/81$ & $- 1/3$ \\
$\psi^\prime(1/3) \alpha^2$ & $5/9$ & $10/9$ & $10/9$ & $10/9$ & $- 1/9$ \\
$\psi^\prime(1/3) \alpha^3$ & $1/9$ & $2/9$ & $2/9$ & $2/9$ & $0$ \\
$\psi^\prime(1/3) \pi^2 $ & $- 8/27$ & $64/81$ & $16/27$ & $32/81$ & $16/27$ \\
$(\psi^\prime(1/3))^2$ & $- 2/9$ & $- 16/27$ & $- 4/9$ & $- 8/27$ & $- 4/9$ \\
$\psi^{\prime\prime\prime}(1/3)$ & $1/108$ & $0$ & $1/54$ & $1/27$ & $2/27$ \\
$\psi^{\prime\prime\prime}(1/3) \alpha$ & $- 1/54$ & $- 1/27$ & $- 1/27$ & $- 1/27$ & $1/54$ \\
$\pi^3 \alpha/\sqrt{3}$ & $29/648$ & $0$ & $29/324$ & $29/162$ & $- 29/972$ \\
$\pi^3/\sqrt{3}$ & $- 145/1944$ & $0$ & $- 145/972$ & $- 145/486$ & $- 493/972$ \\
$\pi \ln(3) \alpha/\sqrt{3}$ & $1/2$ & $0$ & $1$ & $2$ & $- 1/3$ \\
$\pi \ln(3)/\sqrt{3}$ & $- 5/6$ & $0$ & $- 5/3$ & $- 10/3$ & $- 17/3$ \\
$\pi (\ln(3))^2 \alpha/\sqrt{3}$ & $- 1/24$ & $0$ & $- 1/12$ & $- 1/6$ & $1/36$ \\
$\pi (\ln(3))^2/\sqrt{3}$ & $5/72$ & $0$ & $5/36$ & $5/18$ & $17/36$ \\
\hline
\end{tabular}
\end{center}
\begin{center}
{Table $20$. Coefficients of $C_F C_A$ for two loop RI${}^\prime$/SMOM 
$\partial W_2$ amplitudes.}
\end{center}
\end{table}}

\clearpage

{\begin{table}[ht]
\begin{center}
\begin{tabular}{|c||r|r|r|r|}
\hline
$a^{(23)}_n$ & $c^{W_2,(23)}_{(3)\,n}$ & $c^{W_2,(23)}_{(4)\,n}$ 
& $c^{W_2,(23)}_{(5)\,n}$ & $c^{W_2,(23)}_{(6)\,n}$ \\
\hline
$1$ & $718/81$ & $- 443/162$ & $- 683/81$ & $737/81$ \\
$\pi^2 \alpha$ & $1024/729$ & $- 4336/729$ & $- 10376/729$ & $2960/6561$ \\
$\pi^4 \alpha$ & $928/6561$ & $704/6561$ & $1404/6561$ & $608/6561$ \\
$\zeta(3) \alpha$ & $- 8/3$ & $8/3$ & $8$ & $- 8/3$ \\
$\Sigma \alpha$ & $8/9$ & $16/9$ & $32/9$ & $- 8/9$ \\
$\alpha$ & $- 188/81$ & $- 125/162$ & $238/81$ & $- 454/81$ \\
$\pi^2 \alpha^2$ & $116/243$ & $220$/243 & $380/243$ & $- 140/243$ \\
$\pi^4 \alpha^2$ & $128/2187$ & $256/2187$ & $512/2187$ & $- 128/2187$ \\
$\alpha^2$ & $26/27$ & $89/54$ & $68/27$ & $- 32/27$ \\
$\pi^2$ & $- 30452/2187$ & $31286/2187$ & $52576/2187$ & $- 26440/2187$ \\
$\pi^4$ & $- 24064/19683$ & $8776/19683$ & $18656/19683$ & $- 17360/19683$ \\
$\zeta(3)$ & $- 136/3$ & $12$ & $32$ & $- 112/3$ \\
$\Sigma$ & $8/27$ & $- 32/27$ & $- 40/27$ & $40/27$ \\
$s_2(\pi/6)$ & $320$ & $- 144$ & $- 288$ & $224$ \\
$s_2(\pi/6) \alpha$ & $- 32$ & $32$ & $96$ & $- 32$ \\
$s_2(\pi/2)$ & $- 640$ & $288$ & $576$ & $- 448$ \\
$s_2(\pi/2) \alpha$ & $64$ & $- 64$ & $- 192$ & $64$ \\
$s_3(\pi/6)$ & $- 1600/3$ & $240$ & $480$ & $- 1120/3$ \\
$s_3(\pi/6) \alpha$ & $- 1600/3$ & $- 160/3$ & $- 160$ & $160/3$ \\
$s_3(\pi/2)$ & $1280/3$ & $- 192$ & $- 384$ & $896/3$ \\
$s_3(\pi/2) \alpha$ & $- 128/3$ & $128/3$ & $128$ & $- 128/3$ \\
$\psi^\prime(1/3)$ & $15226/729$ & $- 15643/729$ & $- 26288/729$ & $13220/729$ \\
$\psi^\prime(1/3) \pi^2 \alpha$ & $- 64/2187$ & $1024/2187$ & $2048/2187$ & $- 1472/2187$ \\
$\psi^\prime(1/3) \alpha$ & $- 512/243$ & $2186/243$ & $5188/143$ & $- 1480/243$ \\
$\psi^\prime(1/3) \pi^2 \alpha^2 $ & $- 128/729$ & $- 256/729$ & $- 512/729$ & $128/729$ \\
$\psi^\prime(1/3) \alpha^2 $ & $- 58/81$ & $- 110/81$ & $- 190/81$ & $70/81$ \\
$\psi^\prime(1/3) \pi^2 $ & $- 8768/6561$ & $- 3808/6561$ & $- 2240/6561$ & $- 2944/6561$ \\
$(\psi^\prime(1/3))^2$ & $2192/2187$ & $952/2187$ & $560/2187$ & $736/2187$ \\
$(\psi^\prime(1/3))^2 \alpha$ & $16/729$ & $- 256/729$ & $- 512/729$ & $368/729$ \\
$(\psi^\prime(1/3))^2 \alpha^2$ & $32/243$ & $64/243$ & $128/243$ & $- 32/243$ \\
$\psi^{\prime\prime\prime}(1/3)$ & $152/243$ & $- 23/243$ & $- 76/243$ & $94/243$ \\
$\psi^{\prime\prime\prime}(1/3) \alpha$ & $- 4/81$ & $- 8/81$ & $- 16/81$ & $4/81$ \\
$\pi^3 \alpha/\sqrt{3}$ & $58/243$ & $- 58/243$ & $- 58/81$ & $58/243$ \\
$\pi^3/\sqrt{3}$ & $- 580/243$ & $29/27$ & $- 58/27$ & $- 406/243$ \\
$\pi \ln(3) \alpha/\sqrt{3}$ & $8/3$ & $- 8/3$ & $- 8$ & $8/3$ \\
$\pi \ln(3)/\sqrt{3}$ & $- 80/3$ & $12$ & $24$ & $- 56/3$ \\
$\pi (\ln(3))^2 \alpha/\sqrt{3}$ & $- 2/9$ & $2/9$ & $2/3$ & $- 2/9$ \\
$\pi (\ln(3))^2/\sqrt{3}$ & $20/9$ & $- 1$ & $- 2$ & $14/9$ \\
\hline
\end{tabular}
\end{center}
\begin{center}
{Table $21$. Coefficients of $C_F^2$ for two loop RI${}^\prime$/SMOM $W_2$ 
amplitudes.}
\end{center}
\end{table}}

\clearpage 

{\begin{table}[hb]
\begin{center}
\begin{tabular}{|c||r|r|r|r|}
\hline
$a^{(23)}_n$ & $c^{W_2,(23)}_{(7)\,n}$ & $c^{W_2,(23)}_{(8)\,n}$ 
& $c^{W_2,(23)}_{(9)\,n}$ & $c^{W_2,(23)}_{(10)\,n}$ \\
\hline
$1$ & $605/162$ & $- 610/81$ & $83/54$ & $- 83/54$ \\
$\pi^2 \alpha$ & $- 1496/729$ & $- 5272/729$ & $0$ & $- 16/27$ \\
$\pi^4 \alpha$ & $736/6561$ & $- 64/6561$ & $0$ & $- 64/729$ \\
$\zeta(3) \alpha$ & $0$ & $8/3$ & $0$ & $0$ \\
$\Sigma \alpha$ & $8/9$ & $16/9$ & $0$ & $0$ \\
$\alpha$ & $- 523/162$ & $- 244/81$ & $- 7/18$ & $7/18$ \\
$\pi^2 \alpha^2$ & $68/243$ & $220/243$ & $0$ & $-4/27$ \\
$\pi^4 \alpha^2$ & $128/2187$ & $256/2187$ & $0$ & $0$ \\
$\alpha^2$ & $19/54$ & $46/27$ & $1/6$ & $- 1/6$ \\
$\pi^2$ & $- 8930/2187$ & $49028/2187$ & $566/81$ & $1030/81$ \\
$\pi^4$ & $- 17632/19683$ & $5056/19683$ & $872/2187$ & $- 128/2187$ \\
$\zeta(3)$ & $- 92/3$ & $40/3$ & $28/3$ & $- 4/3$ \\
$\Sigma$ & $- 4/27$ & $- 80/27$ & $0$ & $- 4/3$ \\
$s_2(\pi/6)$ & $160$ & $- 224$ & $- 80$ & $128$ \\
$s_2(\pi/6) \alpha$ & $0$ & $32$ & $0$ & $0$ \\
$s_2(\pi/2)$ & $- 320$ & $448$ & $160$ & $320/3$ \\
$s_2(\pi/2) \alpha$ & $0$ & $- 64$ & $0$ & $0$ \\
$s_3(\pi/6)$ & $- 800/3$ & $1120/3$ & $400/3$ & $- 256/3$ \\
$s_3(\pi/6) \alpha$ & $0$ & $- 160/3$ & $0$ & $0$ \\
$s_3(\pi/2)$ & $896/3$ & $- 896/3$ & $- 320/3$ & $- 515/27$ \\
$s_3(\pi/2) \alpha$ & $0$ & $128/3$ & $0$ & $0$ \\
$\psi^\prime(1/3)$ & $4465/729$ & $- 24514/729$ & $- 283/27$ & $- 515/27$ \\
$\psi^\prime(1/3) \pi^2 \alpha$ & $128/2187$ & $1792/2187$ & $0$ & $64/243$ \\
$\psi^\prime(1/3) \alpha$ & $748/243$ & $2636/243$ & $0$ & $8/9$ \\
$\psi^\prime(1/3) \pi^2 \alpha^2 $ & $- 128/729$ & $- 256/729$ & $0$ & $0$ \\
$\psi^\prime(1/3) \alpha^2 $ & $- 34/81$ & $- 110/81$ & $0$ & $2/9$ \\
$\psi^\prime(1/3) \pi^2 $ & $- 4821/6561$ & $- 3328/66561$ & $- 224/729$ & $- 736/729$ \\
$(\psi^\prime(1/3))^2$ & $1208/2187$ & $832/2187$ & $56/243$ & $184/243$ \\
$(\psi^\prime(1/3))^2 \alpha$ & $- 32/729$ & $- 448/729$ & $0$ & $- 16/81$ \\
$(\psi^\prime(1/3))^2 \alpha^2$ & $32/243$ & $64/243$ & $0$ & $0$ \\
$\psi^{\prime\prime\prime}(1/3)$ & $104/243$ & $- 8/243$ & $- 1/9$ & $4/27$ \\
$\psi^{\prime\prime\prime}(1/3) \alpha$ & $- 4/81$ & $- 8/81$ & $0$ & $0$ \\
$\pi^3 \alpha/\sqrt{3}$ & $0$ & $- 58/243$ & $0$ & $0$ \\
$\pi^3/\sqrt{3}$ & $- 290/243$ & $406/243$ & $145/243$ & $116/243$ \\
$\pi \ln(3) \alpha/\sqrt{3}$ & $0$ & $- 8/3$ & $0$ & $0$ \\
$\pi \ln(3)/\sqrt{3}$ & $- 40/3$ & $56/3$ & $20/3$ & $16/3$ \\
$\pi (\ln(3))^2 \alpha/\sqrt{3}$ & $0$ & $2/9$ & $0$ & $0$ \\
$\pi (\ln(3))^2/\sqrt{3}$ & $10/9$ & $- 14/9$ & $- 5/9$ & $- 4/9$ \\
\hline
\end{tabular}
\end{center}
\begin{center}
{Table $22$. Coefficients of $C_F^2$ for two loop RI${}^\prime$/SMOM $W_2$ 
amplitudes continued.}
\end{center}
\end{table}}

\clearpage 
{\begin{table}[ht]
\begin{center}
\begin{tabular}{|c||r|r|r|r|r|}
\hline
$a^{(23)}_n$ & $c^{\partial W_2,(23)}_{(1)\,n}$ & 
$c^{\partial W_2,(23)}_{(3)\,n}$ & $c^{\partial W_2,(23)}_{(4)\,n}$ & 
$c^{\partial W_2,(23)}_{(5)\,n}$ & $c^{\partial W_2,(23)}_{(9)\,n}$ \\
\hline
$1$ & $- 7/2$ & $- 28/3$ & $- 7$ & $- 14/3$ & $0$ \\
$\pi^2 \alpha$ & $- 68/27$ & $- 56/27$ & $- 136/27$ & $- 8$ & $- 8/27$ \\
$\pi^4 \alpha$ & $16/81$ & $32/81$ & $32/81$ & $32/81$ & $0$ \\
$\zeta(3) \alpha$ & $4/3$ & $0$ & $0$ & $0$ & $0$ \\
$\Sigma \alpha$ & $4/3$ & $8/3$ & $8/3$ & $8/3$ & $0$ \\
$\alpha$ & $2$ & $16/3$ & $4$ & $8/3$ & $0$ \\
$\pi^2 \alpha^2$ & $0$ & $0$ & $0$ & $0$ & $- 4/27$ \\
$\pi^4 \alpha^2$ & $0$ & $0$ & $0$ & $0$ & $0$ \\
$\alpha^2$ & $0$ & $0$ & $0$ & $0$ & $0$ \\
$\pi^2$ & $122/27$ & $176/27$ & $244/27$ & $104/9$ & $172/9$ \\
$\pi^4$ & $- 20/81$ & $- 256/243$ & $- 40/81$ & $16/243$ & $8/27$ \\
$\zeta(3)$ & $- 28/3$ & $- 32$ & $- 56/3$ & $- 16/3$ & $8$ \\
$\Sigma$ & $- 2/3$ & $- 8/3$ & $- 4/3$ & $0$ & $- 4/3$ \\
$s_2(\pi/6)$ & $8$ & $96$ & $16$ & $- 64$ & $- 144$ \\
$s_2(\pi/6) \alpha$ & $16$ & $0$ & $32$ & $64$ & $0$ \\
$s_2(\pi/2)$ & $- 16$ & $- 192$ & $- 32$ & $128$ & $288$ \\
$s_2(\pi/2) \alpha$ & $- 32$ & $0$ & $- 64$ & $- 128$ & $0$ \\
$s_3(\pi/6)$ & $- 40/3$ & $- 160$ & $- 80/3$ & $320/3$ & $240$ \\
$s_3(\pi/6) \alpha$ & $- 80/3$ & $0$ & $- 160/3$ & $- 320/3$ & $0$ \\
$s_3(\pi/2)$ & $32/3$ & $128$ & $64/3$ & $- 256/3$ & $192$ \\
$s_3(\pi/2) \alpha$ & $64/3$ & $0$ & $128/3$ & $256/3$ & $0$ \\
$\psi^\prime(1/3)$ & $- 61/9$ & $- 88/9$ & $- 122/9$ & $- 52/3$ & $- 86/3$ \\
$\psi^\prime(1/3) \pi^2 \alpha$ & $0$ & $0$ & $0$ & $0$ & $0$ \\
$\psi^\prime(1/3) \alpha$ & $34/9$ & $28/9$ & $68/9$ & $12$ & $4/9$ \\
$\psi^\prime(1/3) \pi^2 \alpha^2 $ & $0$ & $0$ & $0$ & $0$ & $0$ \\
$\psi^\prime(1/3) \alpha^2 $ & $0$ & $0$ & $0$ & $0$ & $- 32/27$ \\
$\psi^\prime(1/3) \pi^2 $ & $- 16/27$ & $- 128/81$ & $- 32/27$ & $- 64/81$ & $- 32/27$ \\
$(\psi^\prime(1/3))^2$ & $4/9$ & $32/27$ & $8/9$ & $16/27$ & $8/9$ \\
$(\psi^\prime(1/3))^2 \alpha$ & $- 2/27$ & $- 4/27$ & $0$ & $0$ & $0$ \\
$(\psi^\prime(1/3))^2 \alpha^2$ & $0$ & $32/27$ & $8/9$ & $0$ & $0$ \\
$\psi^{\prime\prime\prime}(1/3)$ & $4/9$ & $16/27$ & $1/3$ & $2/27$ & $1/27$ \\
$\psi^{\prime\prime\prime}(1/3) \alpha$ & $- 2/27$ & $- 4/27$ & $- 4/27$ & $- 4/27$ & $0$ \\
$\pi^3 \alpha/\sqrt{3}$ & $- 29/243$ & $0$ & $- 58/243$ & $- 116/243$ & $0$ \\
$\pi^3/\sqrt{3}$ & $- 29/486$ & $- 58/81$ & $- 29/243$ & $116/243$ & $29/27$ \\
$\pi \ln(3) \alpha/\sqrt{3}$ & $- 4/3$ & $0$ & $- 8/3$ & $- 16/3$ & $0$ \\
$\pi \ln(3)/\sqrt{3}$ & $- 2/3$ & $- 8$ & $- 4/3$ & $16/3$ & $12$ \\
$\pi (\ln(3))^2 \alpha/\sqrt{3}$ & $1/9$ & $0$ & $2/9$ & $4/9$ & $0$ \\
$\pi (\ln(3))^2/\sqrt{3}$ & $1/18$ & $2/3$ & $1/9$ & $- 4/9$ & $- 1$ \\
\hline
\end{tabular}
\end{center}
\begin{center}
{Table $23$. Coefficients of $C_F^2$ for two loop RI${}^\prime$/SMOM 
$\partial W_2$ amplitudes.}
\end{center}
\end{table}}

\end{document}